\documentclass[useAMS,usenatbib]{mn2e}
\usepackage{graphicx}
\usepackage[figuresright]{rotating}

\def\halpha{H$\alpha$}
\def\teff{$ T_\mathrm{eff}$}

\def\mbol{$M_\mathrm{bol}$}

\title[An {\it XMM-Newton} view of NGC~6231. III.]{An {\it XMM-Newton} view of the young open cluster NGC~6231\thanks{Based on observations with {\it XMM-Newton}, an ESA science mission with instruments and contributions directly funded by ESA member states and the USA (NASA).} III. Optically faint X-ray sources}
\author[H.~Sana et al.~]
  {H. Sana$^{1,2}$\thanks{E-mail: hsana@eso.org}, 
   G. Rauw$^1$\thanks{FNRS Research Associate (Belgium)},
   H. Sung$^3$,
   E. Gosset$^1\ddagger$\ and J.-M. Vreux$^1$\\
$^1$ Institut d'Astrophysique et de G\'eophysique, University of Li\`ege, All\'ee du 6 Ao\^ut 17, B\^at. B5c, B-4000 Li\`ege, Belgium\\
$^2$ European Southern Observatory, Alonso de Cordova 3107, Vitacura, Casilla 19001, Santiago 19, Chile\\
$^3$ Department of Astronomy and Space Science, Sejong University, Kunja-dong 98, Kwangjin-gu, Seoul 143-747, Korea}

\begin{document}

\date{Accepted 1988 December 15. Received 1988 December 14; in original form 1988 October 11}

\pagerange{\pageref{firstpage}--\pageref{lastpage}} \pubyear{2002}

\maketitle

\label{firstpage}

\begin{abstract}
We discuss the properties of the X-ray sources with faint optical counterparts in the very young open cluster NGC\,6231. From their positions in the H-R diagram, we find that the bulk of these objects probably consists of low-mass pre-main sequence stars with masses in the range 0.3 to 3.0\,M$_{\odot}$. The age distribution of these objects indicates that low-mass star formation in NGC\,6231 started more than 10\,Myr ago and culminated in a starburst-like event about 1 to 4\,Myr ago when the bulk of the low-mass PMS stars as well as the massive cluster members formed. We find no evidence for a spatial age gradient that could point towards a sequential star formation process. Only a few X-ray sources have counterparts with a reddening exceeding the average value of the cluster or with infrared colours indicating the presence of a moderate near-IR excess. The X-ray spectra of the brightest PMS sources are best fitted by rather hard thermal plasma models and a significant fraction of these sources display flares in their light curve. The X-ray brightest flaring sources have decay times between 2 and 16\,ks. The X-ray selected PMS stars in NGC\,6231 have $\log{L_{\rm X}/L_{\rm bol}}$ values that increase strongly with decreasing bolometric luminosity and can reach a saturation level ($\log{L_{\rm X}/L_{\rm bol}} \sim -2.4$) for non-flaring sources and even more extreme values during flares.
\end{abstract}


\section{Introduction}
The formation of massive stars is currently one of the key questions in stellar astrophysics. Various scenarios have been proposed; accretion from a circumstellar envelope (e.g.\ Shu et al.\ \citeyear{SAL}, Behrend \& Maeder \citeyear{BM1}), accretion from a turbulent molecular core (McKee \& Tan \citeyear{MT}) or a mixture of competitive accretion and collisions of lower mass protostars in the core of dense stellar clusters (e.g.\ Bonnell et al.\ \citeyear{BBZ}, Bonnell \& Bate \citeyear{BB2}) being the most popular ones. Since the bulk of the massive stars are found in open clusters, some clues on this question can probably be obtained from a study of the star formation processes within young open clusters. Beside their massive star population, such clusters usually harbour a wealth of lower mass objects with the least massive ones that have not yet reached the zero age main sequence. An important issue is the relationship between the high-mass cluster members and these low-mass objects. While it has sometimes been suggested that the massive stars trigger the formation of the lower mass objects, in several cases, the lower mass pre-main sequence stars are found to be significantly older than the massive cluster members (e.g.\ Damiani et al.\ \citeyear{Damiani}, Prisinzano et al.\ \citeyear{Prisinzano}). Herbig (\citeyear{herbig}) accordingly suggested that the formation of early-type stars could stop the formation process of lower mass stars in a cluster by dispersing the ambient gas. Other authors (e.g.\ Sung et al.\  \citeyear{SBL97}) however suggested that the observed age discrepancy might result from the fact that the PMS evolutionary tracks actually overestimate the {\it real} age of the protostars. Obviously, investigating the properties of lower mass pre-main sequence (PMS) stars in very young open clusters can help us to understand the feedback of massive stars on their environment and can thereby shed light on the way the most massive objects have formed.
  
Classical T\,Tauri PMS stars display emission in the Balmer lines and can thus be identified through photometric observations with an H$\alpha$ narrow band filter. Most classical T\,Tauri stars (cTTs) also display near-infrared excesses that are attributed to heated dust in a disk-like accretion structure (e.g.\ Meyer et al.\ \citeyear{meyer}). However, another category of PMS objects, the so-called weak-line T\,Tauri stars (wTTs), display only weak optical line emission and do not show significant near-IR excesses. An interesting property that can help us to identify these PMS stars is their relatively strong X-ray emission with $\log{L_{\rm X}/L_{\rm bol}}$ reaching values as large as $-3$ (e.g.\ Neuh\"auser \citeyear{Neu}, Feigelson \& Montmerle \citeyear{FM}). X-ray observations can therefore enable the selection of PMS stars, such as wTTs, that would neither be detected through H$\alpha$ nor through near-IR photometry (e.g.\ Damiani et al.\ \citeyear{Damiani}). The high sensitivity of {\it XMM-Newton} and the exceptional spatial resolution of {\it Chandra} have already been used to study some rich young open clusters in the X-ray domain. For instance, NGC\,6530 at the core of the Lagoon Nebula was found to harbour both classical and weak-line T\,Tauri stars with masses between 0.5 and 2.0\,M$_{\odot}$, but only a few of the X-ray selected PMS candidates were found to be cTTs (Rauw et al.\ \citeyear{M8}, Damiani et al.\ \citeyear{Damiani}).\footnote{We note here that the luminosities derived by Rauw et al.\ (\citeyear{M8}) for the PMS objects in NGC\,6530 are too low, leading to an overestimate of their age. The reason for this error is that the $V - I_C$ colour index quoted by Sung et al.\ (\citeyear{SCB}) are actually given in the Cousins photometric system while the {\tt Vizier} database erroneously indicates that these are expressed in the Johnson system.} NGC\,6383, on the other hand, does not contain any known cTTs, but our {\it XMM-Newton} observations revealed a number of weak X-ray sources in this cluster that are most likely associated with wTTs (Rauw et al.\ \citeyear{ngc6383}). A preliminary analysis of the optical properties of these sources suggested that they are in fact older than the massive binary HD\,159176 (O7\,V + O7\,V) in the centre of the cluster. 

NGC\,6231 - at the core of the Sco\,OB1 association - is another very young open cluster, rich in massive O-type stars.  Using $U\!BV\!(I)_\mathrm{C}\!H\alpha$ photometry, Sung et al.\ (\citeyear{SBL98}, hereafter SBL98) inferred a mean cluster reddening of $\overline{E(B - V)} = 0.466 \pm 0.054$ and a distance modulus of $11.0 \pm 0.07$. These authors noted that the reddening law towards NGC\,6231 could be somewhat peculiar with $R_V = 2.45\,E(V-I_C)/E(B-V) = 3.3 \pm 0.1$. SBL98 found only 12 PMS objects (and 7 PMS candidates) brighter than $V = 17$ displaying H$\alpha$ emission. However, as pointed out above, there could be a number of PMS stars without H$\alpha$ emission; but with optical photometry only, these would be very difficult to distinguish from field stars lying to the right of the cluster main-sequence. In this paper, we use the catalogue of {\it XMM-Newton} sources in NGC\,6231 presented by Sana et al.\ (\citeyear{paperI}, hereafter Paper~I) to investigate the properties of X-ray sources with faint optical counterparts in this cluster. The X-ray properties of the early-type stars in NGC\,6231 were discussed in a separate paper (Sana et al.\ \citeyear{paperII}, Paper~II). 

\section{Optical counterparts }\label{optcoun}
Adopting a 3\arcsec\ correlation radius, 446 of the X-ray sources in the NGC\,6231 field have an optical counterpart in the catalogue of Sung et al.\ (\citeyear{SBL98}) extended down to $V = 21$ (Sung H., unpublished).  We note that our combined EPIC field of view extends over a wider area than the actual field investigated by SBL98 and some sources outside the SBL98 area actually have counterparts in other optical/near-IR catalogues (see Paper~I). Nevertheless, throughout this paper, we will restrict ourselves to the sources with counterparts in the extended SBL98 catalogue. The reasons are, on the one hand, that the USNO and GSC photometries are less accurate than the results of SBL98. On the other hand, the new photometry acquired by one of us (H.\ Sung) over a larger field than initially investigated by SBL98 (referred to as SSB06 in Paper~I) does not include observations with the H$\alpha$ filter. 

\begin{figure}
\begin{center}
\resizebox{8cm}{7.9cm}{\includegraphics{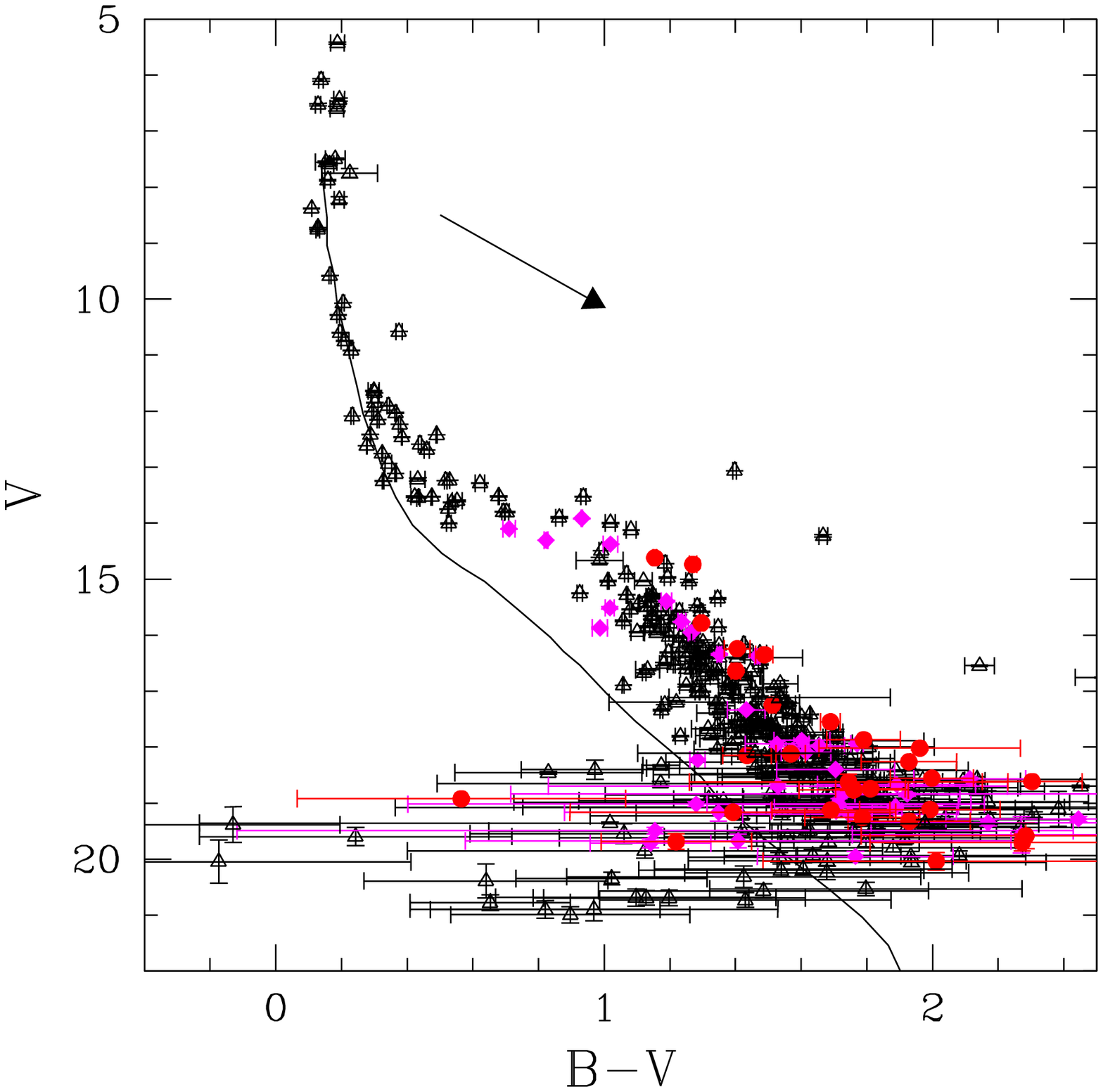}}
\resizebox{8cm}{7.9cm}{\includegraphics{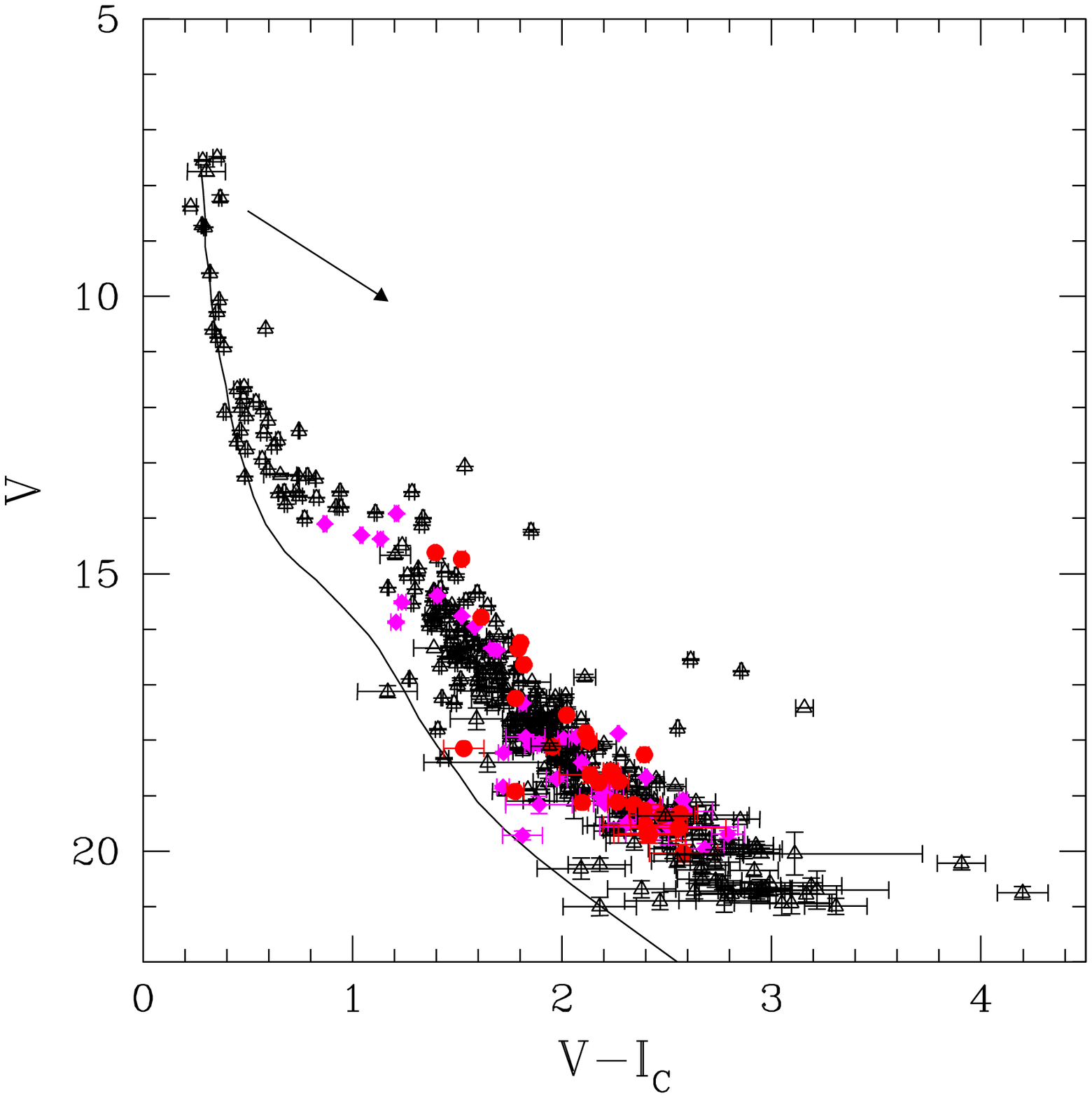}}
\end{center}
\caption{Colour-magnitude diagrams of the X-ray sources with an optical counterpart in the SBL98 catalogue. Filled dots, filled diamonds and open triangles indicate respectively H$\alpha$ emitting stars, H$\alpha$ candidates and stars with no evidence for emission (see Fig.\,\ref{ha}). The reddening vector with $R_V = 3.3$ and $E(V-I_C)/E(B-V) = 1.365$ (Sung et al.\ \citeyear{SBL98}) is indicated and the solid line shows the ZAMS relation taken from Sung \& Bessell (\citeyear{SB99}), shifted by a distance modulus DM = 11.07 and reddened with $E(B-V) = 0.466$ (Sung et al.\ \citeyear{SBL98}). } \label{coulmag}
\end{figure}

The $V$ vs.\ $B - V$ and $V$ vs.\ $V - I_C$ colour-magnitude diagrams of the optical counterparts are shown in Fig.\,\ref{coulmag}. The bulk of the objects are found to be rather faint stars mostly located to the right of the main-sequence. At $V=18$, the photometric error on most colour indices reaches about 0.1~mag. Errors on the $B-V$ index tend however to be larger (Fig.~\ref{coulmag}, upper panel) because of the lower quality of the $B$ photometry in SBL98.  This fortunately does not affect our work as we focus on the $V$, $V-I_\mathrm{C}$ and $R - H\alpha$ measurements.

\begin{figure}
\begin{center}
\resizebox{8cm}{7.9cm}{\includegraphics{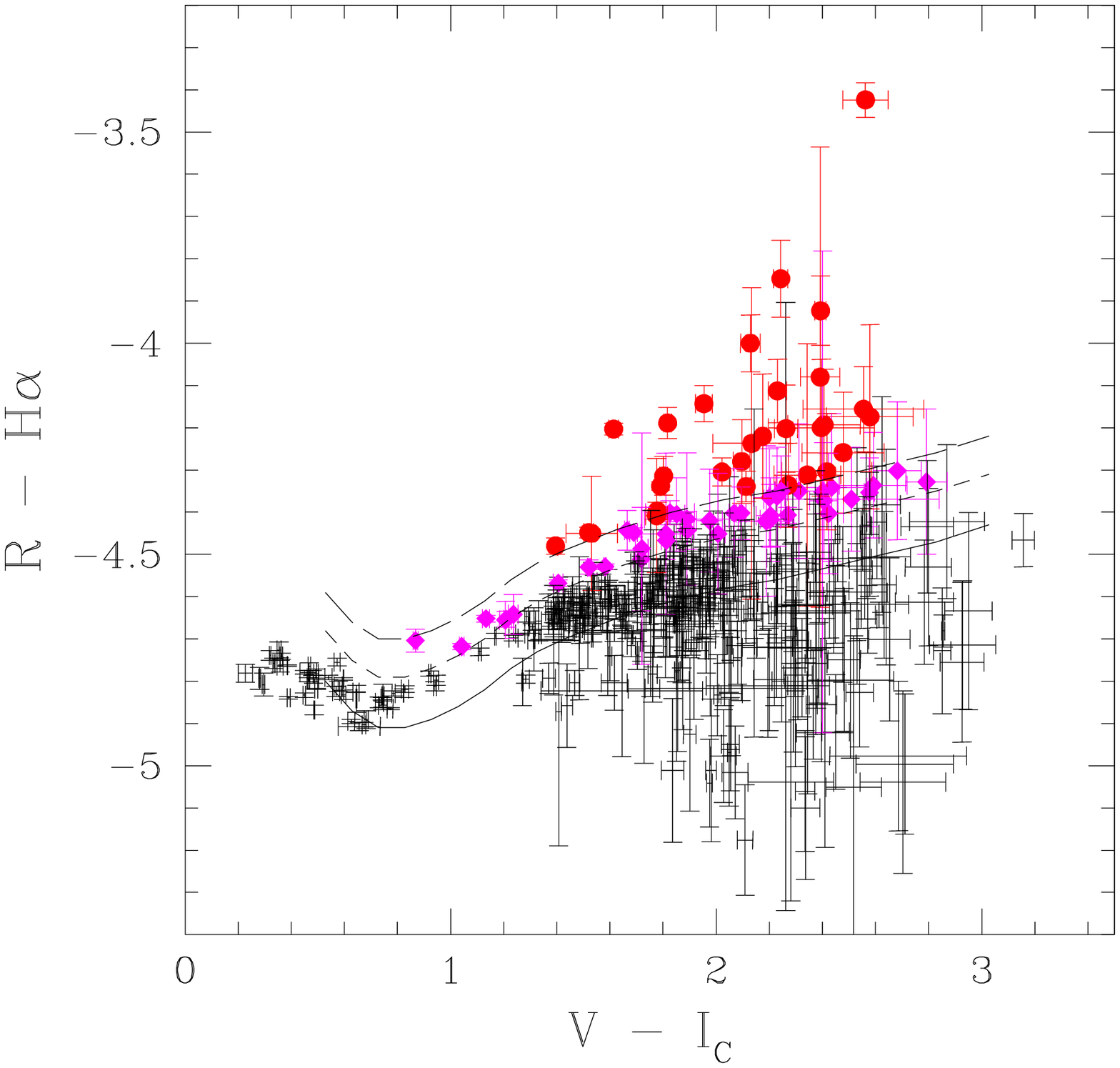}}
\resizebox{8cm}{7.9cm}{\includegraphics{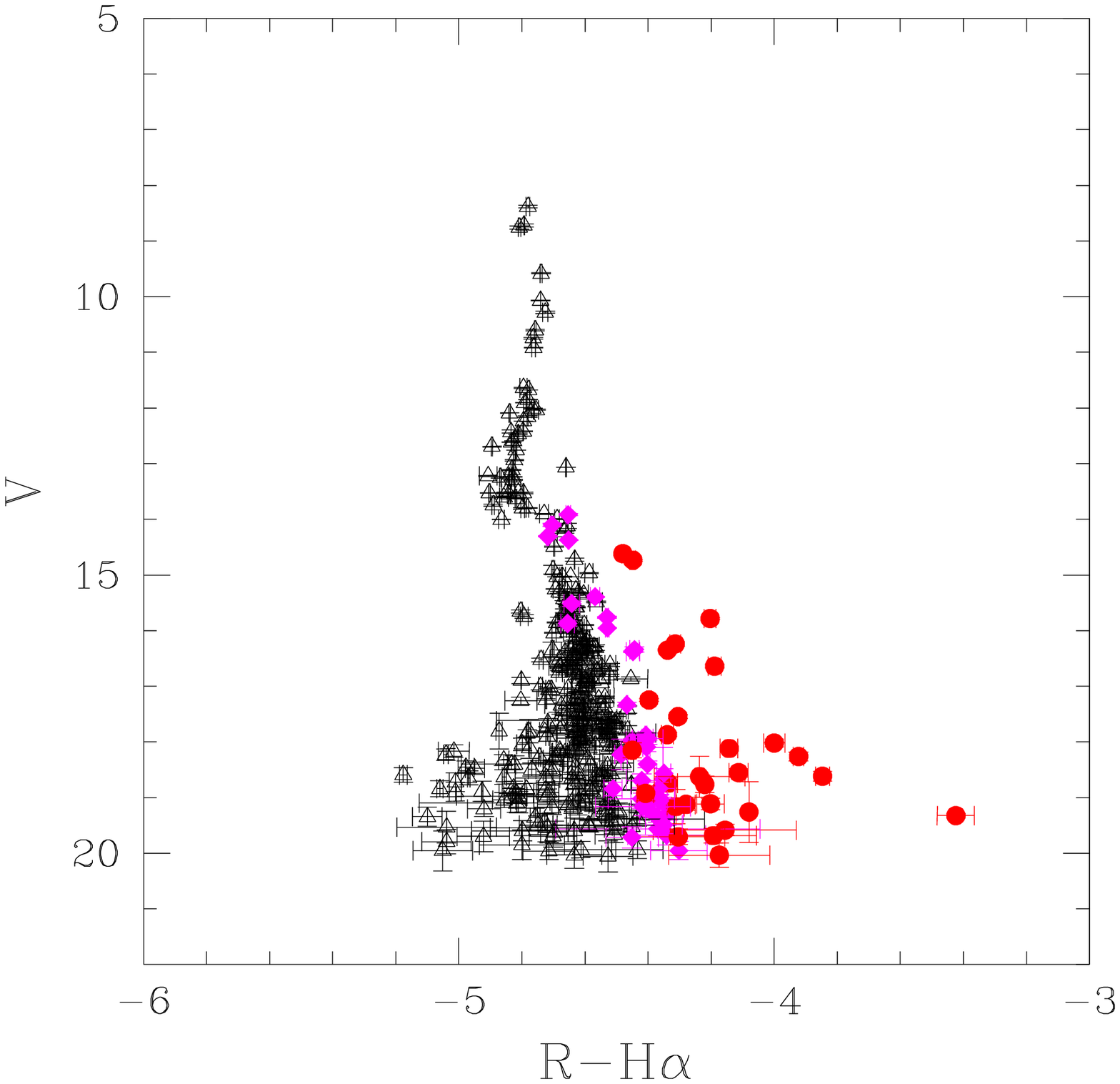}}
\end{center}
\caption{ {\bf Top  panel:} $R - H\alpha$ as a function of $V - I_C$ for the SBL98 optical counterparts.  The solid line yields the $R - H\alpha$ versus $V - I_C$ relation for main sequence stars taken from Sung et al.\ (\citeyear{SBL97}) and reddened with the average reddening of NGC\,6231. The short- and long-dashed lines yield respectively the thresholds for H$\alpha$ emission candidates and H$\alpha$ emitters (see text). {\bf Bottom panel:} $V$-magnitude as a function of $R - H\alpha$ for the SBL98 optical counterparts of our EPIC sources. The symbols have the same meaning as in Fig.\,\ref{coulmag}.} \label{ha}
\end{figure}

Among the 446 counterparts in the extended version of the SBL98 catalogue, 392 sources  have been observed in the $V$, $I_\mathrm{C}$ and H$\alpha$ bands. This allowed the authors to compute a pseudo continuum (labelled $R$) near H$\alpha$, as the mean of the $V$ and $I_\mathrm{C}$ magnitudes. To identify H$\alpha$ emitting stars among these X-ray selected objects, one can then compare the $R - H\alpha$ index to its value for main sequence stars. For the latter, we use the relation between the $R - H\alpha$ and $V - I_C$ indices for main sequence stars proposed by Sung et al.\, (\citeyear{SBL97}) and we account for the effect of cluster reddening on the $V - I_C$ colour index (see below). Following Sung et al.\ (\citeyear{SBL97}), we consider that a star displays H$\alpha$ emission if $\Delta(R - H\alpha) = (R - H\alpha) - (R - H\alpha)_{\rm MS} \geq 0.21$\,mag, whereas a star will be considered an H$\alpha$ emission candidate if $0.12 \leq \Delta(R - H\alpha) \leq 0.21$. 19 objects of our sample have $V-I < 0.50$  and one has $V-I > 3.0$ . Hence, they do not fall in the range covered by the $(R - H\alpha)_{\rm MS}$ relation of Sung et al. (\citeyear{SBL97}). From a visual interpolation (see Fig. 2, upper panel), one can still conclude that these are definitely not \halpha\ emitters. In this way, we find that about 18\% (70 out of 392) of the X-ray selected objects are either confirmed or potential H$\alpha$ emission objects. This large number of H$\alpha$ emission candidates compared to the original results of SBL98 is mainly due to the fact that the photometry used here extends down to much fainter magnitudes (Fig.\,\ref{ha}, lower panel).

\begin{figure}
\begin{center}
\resizebox{8cm}{7.9cm}{\includegraphics{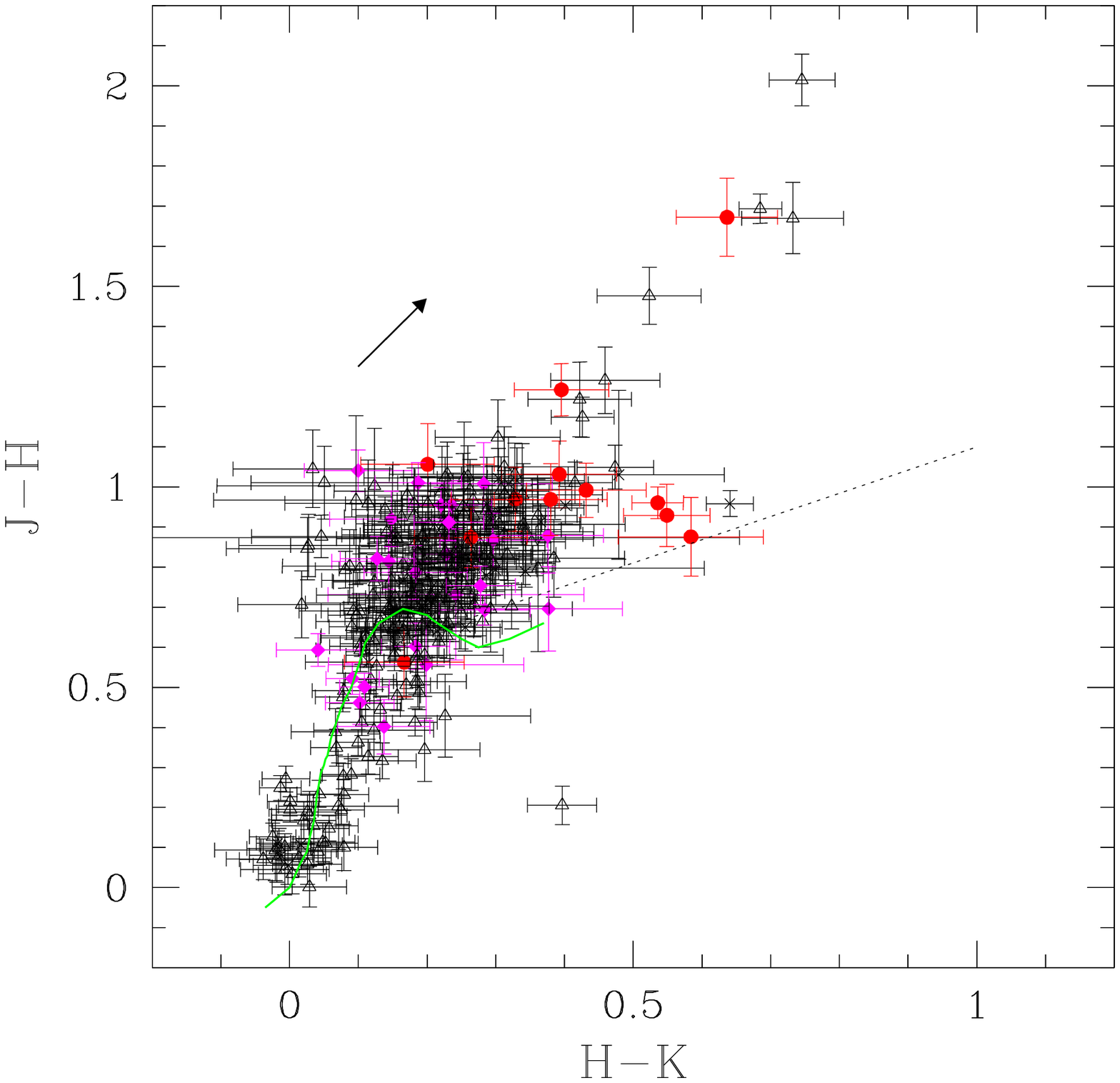}}
\end{center}
\caption{$JHK$ colour-colour diagram of the 2MASS counterparts, with good quality near-IR photometry, of the X-ray sources in the EPIC field of view around NGC\,6231. The heavy solid line yields the intrinsic colours of main sequence stars following Bessell \& Brett (\citeyear{BB}), whereas the reddening vector is illustrated for $A_V = 1.538$ and adopting the reddening law of Rieke \& Lebofsky (\citeyear{RL}). The crosses stand for 2MASS sources having no counterpart in the SBL98 catalogue (mainly because they fall outside the field of view investigated by Sung et al.\ \citeyear{SBL98}). The other symbols have the same meaning as in Fig.\,\ref{coulmag}. The dotted straight line yields the locus of dereddened colours of classical T\,Tauri stars according to Meyer et al.\ (\citeyear{meyer}).} \label{2mass}
\end{figure}

A total of 384 EPIC sources in the field of view of NGC\,6231 have also a counterpart in the Two Micron All Sky Survey point source catalogue (2MASS, Cutri et al.\ \citeyear{2mass}). Among these, 333 sources have also a counterpart in the SBL98 catalogue. We used the March 2003 update of the colour transformations, initially derived by Carpenter (\citeyear{Carpenter}) and available on the 2MASS website\footnote{\tt http://www.ipac.caltech.edu/2mass/index.html}, to convert the $J - H$ and $H - K_s$ colour indices to the homogenized $JHK$ photometric system introduced by Bessell \& Brett (\citeyear{BB}). Assuming that the 2MASS and SBL98 magnitudes are not affected by photometric variability, we tried to establish the effective temperature and bolometric correction using a dereddened $V - K$ colour index\footnote{For this purpose, we use the reddening law of Rieke \& Lebofsky (\citeyear{RL}).}. Comparing the results with those obtained from the $V - I_C$ index (see below), we find that, for the large majority of the stars, the former technique yields systematically lower temperatures (sometimes by more than 1000\,K) than the latter one. For objects surrounded by circumstellar material, such a discrepancy could reflect the presence of a near-IR excess (that would mimic a lower temperature in the $V - K$ colour index). However, for the bulk of the objects, the most likely explanation for this effect is probably the rather large errors in the 2MASS magnitudes. Note also that the reddening correction over such a wide wavelength range, using the extinction law of Rieke \& Lebofsky (\citeyear{RL}), might simply not be adequate for stars in NGC\,6231 (perhaps as a result of the peculiarities of the extinction law discussed by Sung et al.\ \citeyear{SBL98}). In addition, the 2MASS colour indices and magnitudes of some objects are either subject to large uncertainties or are only upper limits due to non-detections. In total, we find that only 295 out of the 384 2MASS counterparts have quality flags A, B, C or D for the measurements of all three individual near-IR magnitudes. Restricting the $JHK$ colour-colour diagram to these objects, we find that their location is consistent with slightly reddened main-sequence or giant stars (see Fig.\,\ref{2mass}). Only a couple of sources show evidence for a moderate IR excess and only about ten objects display strongly reddened IR colours.

Although a small fraction ($\sim 9$\%, see Paper~I) of the X-ray selected objects might in fact be field stars (either foreground or background) unrelated to NGC\,6231, we have assumed that all stars are located at the distance of the cluster (DM = $11.07\pm0.04$, Paper~I) and are all subject to the same reddening ($A_V = 1.538$, Sung et al.\ \citeyear{SBL98}). We have then built the Hertzsprung-Russell diagram (HRD) of these EPIC sources using effective temperatures and bolometric corrections interpolated from the dereddened $V - I_C$ colour indices for main sequence stars of spectral types B0 to M6 tabulated by Kenyon \& Hartmann (\citeyear{KH}). Eight sources (a subset of those not covered by the \citeyear{SBL97} Sung et al.\ relation) do not fall in this range and are therefore not plotted in the HRD.

\begin{figure}
\begin{center}
\resizebox{8cm}{7.9cm}{\includegraphics{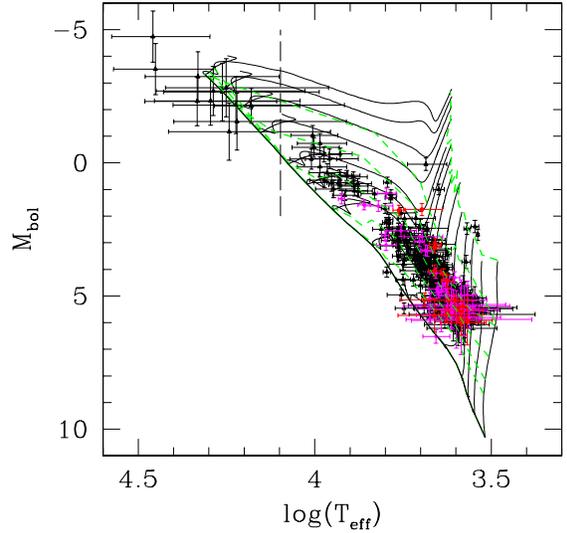}}
\end{center}
\caption{Hertzsprung-Russell diagram of the EPIC sources with optical counterparts in the SBL98 catalogue.  The symbols have the same meaning as in Fig.\,\ref{coulmag}. Evolutionary tracks from Siess et al.\ (\citeyear{siess}) for masses of 0.2, 0.3, 0.4, 0.5, 0.7, 1.0, 1.5, 2.0, 2.5, 3.0, 4.0, 5.0, 6.0 and 7.0\,M$_{\odot}$ are overplotted. The thick solid line shows the ZAMS. The thick dashed line indicate the birth isochrone (i.e. the 0\,Myr isochrone) from Siess et al.  while the other dashed lines correspond to isochrones for ages of 0.5, 1.5, 4.0, 10.0 and 20.0\,Myr. The stars located to the left of the vertical dashed line were not considered in the age distribution of the cluster.}
\label{hrd} 
\end{figure}

Figure\,\ref{hrd} compares the location of the X-ray selected objects to the pre-main sequence evolutionary tracks of Siess et al.\ (\citeyear{siess}) for $Z = 0.02$ and without overshooting (note that these evolutionary models include neither rotation nor accretion). Errors on the effective temperatures and bolometric magnitudes were obtained by error propagation, including contributions from the photometric measurements, the cluster distance, and the average $E(B-V)$ and $E(V-I)$ color excesses obtained from SBL98. Note that the color degeneracy of the hotter stars (i.e. the fact that a small range in color index corresponds to a large range of \teff) implies that the determination of their effective temperature becomes very sensitive to any uncertainty on the $(V-I)_0$ index. Even when the observed $V-I$ data are excellent, the dereddening introduces a fix contribution to the error on  $(V-I)_0$, yielding uncertainties on the effective temperature of several thousands Kelvin (see Fig. \ref{hrd}).  In the following,  the twelve stars with $T_\mathrm{eff}>12500$~K will be ignored while computing the age distribution of the stars in NGC 6231 as they do not provide any constraint on the cluster age.

\begin{figure}
\begin{center}
\resizebox{8cm}{7.7cm}{\includegraphics{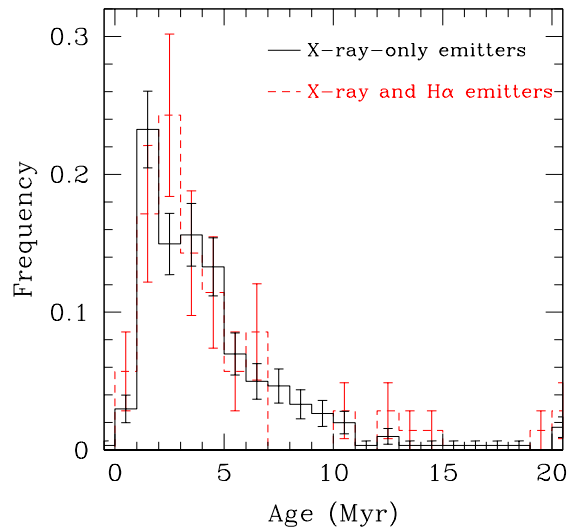}}
\end{center}
\caption{Distribution of the ages of X-ray selected PMS objects as interpolated from the isochrones derived from the Siess et al.\ (\citeyear{siess}) evolutionary tracks. PMS candidates with $\Delta(R - H\alpha) \geq 0.12$ are indicated by the dashed line, whilst those with $\Delta(R - H\alpha) < 0.12$ are indicated by the solid line. The total numbers of objects with $\Delta(R - H\alpha) \geq 0.12$ and $\Delta(R - H\alpha) < 0.12$ are respectively 70 and 302. 1-$\sigma$ Poisson uncertainties have been indicated for both distributions.}\label{ages}
\end{figure}

We find that most of the X-ray selected, optically faint objects ($V \geq 14$) fall between the PMS evolutionary tracks for stars of masses in the range 0.3 to 3.0\,M$_{\odot}$. A comparison with the isochrones further indicates that the majority of these objects should have ages ranging from 1 to 10\,Myr (see Fig.\,\ref{ages}). We estimated the age of the 372 objects by linearly interpolating the isochrones (computed between 0 and 20 Myr with a step of 1Myr) along the \mbol\ axis. The obtained distribution of ages peaks around $\sim$ 1 -- 4\,Myr and decreases towards lower ages (see Fig.~\ref{ages}). Average and quartiles are given in Table \ref{tab: qwart}. The mean and 1-$\sigma$ dispersion are strongly affected by the older age tail and does not  represent neither the centroid nor the width of the peak seen in  the histograms.  Within our uncertainties, no significant difference is found between the different  distributions considered in  Table \ref{tab: qwart}.

We note that Sung et al.\ (\citeyear{SBL98}) compared the H-R diagram of NGC\,6231 to PMS tracks of Bernasconi \& Maeder (\citeyear{BM}) finding an age spread of about 11\,Myr for the low mass objects. While this is similar to the age spread found here, we caution however that a direct comparison of age determinations of PMS stars made with different evolutionary models and different calibrations might be biased (see e.g.\ the discussion in Siess et al.\ \citeyear{siess}). As for NGC\,6530, we do not find a large age difference between those stars displaying H$\alpha$ emission (probable cTTs) and those without (probable wTTs). What we do see, is that H$\alpha$ emission seems restricted to X-ray selected stars with masses below about 2.5\,M$_{\odot}$. This is again reminiscent of the situation in NGC\,6530.

Zinnecker (\citeyear{zinnecker}) noted that the binary frequency among pre-main sequence stars is at least as high as among main sequence stars. Binarity affects the distribution of a coeval population of PMS stars in the H-R diagram by introducing a band shifted upwards from the true isochrone (Siess et al.\ \citeyear{Siess}) leading to an underestimate of the actual age. Palla (\citeyear{Palla}) evaluated the average age discrepancy for a realistic distribution of binary mass ratios, finding that age estimates that do not account for binarity can be off by a factor $ \sim 1.5$. Therefore, binarity should not have too large an impact on our age estimates above. In any case, binarity cannot explain the tail of the age distributions towards older ages (see Fig.\,\ref{ages}).\\

\begin{table}
\caption{Average, dispersion and quartiles (in Myr) of the age distributions of the PMS populations in NGC~6231. $N$ indicates the number of objects considered in each (sub-)population. }
\label{tab: qwart}
\centering
\begin{tabular}{ccccccc}
\hline
Population &  $N$ & mean & 1-$\sigma$  & \multicolumn{3}{c}{Quartiles}\\
           & &      &  dispers. & 1st & 2nd & 3rd \\
\hline
wTTs  & 302 & 4.6 & 3.7 & 1.9 & 3.4 & 4.9 \\
cTTs  &  70 & 4.8 & 4.5 & 1.9 & 2.9 & 3.3 \\
All   & 372 & 4.6 & 3.9 & 1.9 & 3.5 & 5.6 \\
\hline
\end{tabular}
\end{table}

\section{X-ray properties of the faint sources}
\subsection{X-ray spectra}
Spectra were extracted for each source, for each instrument and each observation using the procedure described in Paper~II.
Briefly, we adopted circular extraction regions with a radius corresponding to half the distance to the nearest neighbouring X-ray source. For each source, the background spectrum was obtained over source-free regions chosen according to the source location.  We adopted the redistribution matrix files  provided by the {\it XMM-Newton} SOC, whereas the ancillary response files were built with the appropriate SAS tasks. The spectra were binned to have at least 10 counts per energy channel. In the following, we discuss only the spectra of those sources that have an average, vignetting and exposure-corrected count rate\footnote{We note that the actual (uncorrected) count rates are substantially lower for sources near the edge of the field of view.} above $5 \times 10^{-3}$ or $10^{-2}$\,cts\,s$^{-1}$ for EPIC-MOS or pn respectively and are not associated with OB stars. The X-ray spectra of the OB stars were discussed in a separate paper (Paper~II). We use the numbering scheme of the sources as introduced in Paper~I. 

For sources that display no significant variability, we have analyzed the merged spectra of all observations, while for those sources that experience a flare in their light curve, we rather analyze the spectra of the observation when the flare occurred. The spectra were then modelled by means of rather simple models: absorbed {\tt mekal} optically thin thermal plasma model (Mewe et al.\,\citeyear{mewe}; Kaastra \citeyear{ka}) with one or two temperatures, absorbed power law, absorbed bremsstrahlung model or absorbed blackbody energy distribution (see Tables\,\ref{tbl-1} to \ref{tbl-3}.).

Most of the flaring sources display a spectrum that is best represented by a single temperature {\tt mekal} model with best fit temperatures of k$T \geq 3.4$\,keV. For the majority of the non-flaring bright X-ray sources, a second temperature is needed to model the spectra. In these cases, we find that most of the objects (except sources \#478 and \#535) have moderate hydrogen column densities ($N_{\rm H} \leq 0.36 \times 10^{22}$\,cm$^{-2}$, in reasonable agreement with the average $N_{\rm H}$ estimated from the cluster reddening, see below) and temperatures of order 0.7 and 2.5 -- 3.5\,keV for the soft and hard component respectively. The flaring sources thus appear to have significantly harder spectra than the non-flaring objects. Finally, some of the objects have spectra that are best modelled by a power law spectral energy distribution. Three out of five sources in this group have no optical counterpart (not even in the deeper extension of the SSB06 catalogue).

\begin{table*}
\caption{Best fit parameters of the absorbed single temperature {\tt mekal} models ({\tt wabs*mekal}) fitted to the EPIC spectra of faint X-ray sources in NGC\,6231. The source numbering scheme is adopted from Paper~I. The third and fourth columns indicate the combinations of observations and instruments used in the spectral analysis. Columns $[9]$ and $[10]$ yield the observed and absorption-corrected X-ray flux respectively. The fluxes are expressed in $10^{-14}$\,erg\,cm$^{-2}$\,s$^{-1}$ and are evaluated over the 0.5 -- 10.0\,keV energy band. Col. 11 yields the internal numbering scheme of SBL98. Numbers between brackets are rather taken from the most recent extension of the SSB06 catalogue (see Paper~I) and mostly correspond to sources out of the SBL98 field of investigations. }\label{tbl-1}
\tiny
\begin{center}
\begin{tabular}{c c c c c c c c r r c c c} 
\hline
\multicolumn{2}{c}{Source IDs} & Obs.\ & Inst.\ & $N_{\rm H}$            & k$T$  & $\chi^2_{\nu}$ & d.o.f. & f$_{\rm X}^{\rm obs}$      & f$_{\rm X}^{\rm corr}$ & SBL98 & $V$ & $V-I_C$ \\
XMM\,UJ & Paper~I       &            &        & ($10^{22}$\,cm$^{-2}$) & (keV) &               &        & & & & & \\
$[1]$ & $[2]$ & $[3]$ & $[4]$ & $[5]$ & $[6]$ & $[7]$ & $[8]$ & $[9]$ & $[10]$ & $[11]$ & $[12]$ & $[13]$ \\ 
\hline

165334.5-415858 & \#43 & all & M2 + pn & $0.32^{+.04}_{-.04}$ & $8.96^{+2.22}_{-1.48}$ & 1.05 & 440 & $22.4$ & $26.6$ &  &  & \\
165335.9-413807 & \#48 & \#4 & M2 & $0.16^{+.20}_{-.07}$ & $3.43^{+2.83}_{-1.20}$ & 1.62 & 19 & $22.4$ & $26.5$ & [1987] & [15.89] & [1.47] \\
165348.3-414008 & \#100 & \#3 & M2 + pn & $0.19^{+.06}_{-.05}$ & $4.32^{+0.89}_{-0.69}$ & 0.87 & 79 & $24.2$ & $28.6$ & 4664 & 20.69 & 2.90 \\
165358.6-414643 & \#175 & \#4 & M1 + M2 & $0.15^{+.07}_{-.06}$ & $4.26^{+1.08}_{-0.73}$ & 1.01 & 66 & $37.9$ & $43.5$ & 5433 & 18.27 & 2.08 \\
165410.8-413918 & \#285 & \#1 & EPIC & $0.23^{+.04}_{-.05}$ & $4.44^{+0.78}_{-0.55}$ & 1.19 & 153 & $34.2$ & $41.1$ & [3321] & [17.36] & [2.04] \\
165429.2-415047 & \#442 & \#1 & EPIC & $0.17^{+.05}_{-.05}$ & $4.04^{+0.62}_{-0.54}$ & 1.29 & 118 & $13.9$ & $16.2$ & 7554 & 20.24 & 2.18 \\
165431.9-415107 & \#467 & \#3 & EPIC & $0.27^{+.06}_{-.06}$ & $5.70^{+1.80}_{-0.95}$ & 1.09 & 129 & $21.8$ & $26.2$ & 7757 & 18.12 & 1.95 \\
165432.0-415155 & \#469 & \#6 & EPIC & $0.27^{+.06}_{-.06}$ & $4.17^{+0.76}_{-0.59}$ & 1.20 & 105 & $18.1$ & $22.6$ & 7772 & 17.46 & 1.87 \\
165456.8-420156 & \#583 & all & pn & $0.11^{+.15}_{-.11}$ & $5.58^{+10.61}_{-2.23}$ & 0.89 & 27 & $3.2$ & $3.4$ & [25121] & [18.08] & [2.21] \\
\hline
\end{tabular}
\end{center}
\end{table*}

\begin{table*}
\caption{Same as Table\,\ref{tbl-1}, but for the absorbed power law models ({\tt wabs*power}).}\label{tbl-2}
\tiny
\begin{center}
\begin{tabular}{c c c c c c c c r r c c c} 
\hline
\multicolumn{2}{c}{Source IDs} & Obs.\ & Inst.\ & $N_{\rm H}$            & $\Gamma$  & $\chi^2_{\nu}$ & d.o.f. & f$_{\rm X}^{\rm obs}$      & f$_{\rm X}^{\rm corr}$ & SBL98 & $V$ & $V-I_C$ \\
XMM\,UJ & Paper~I       &            &        & ($10^{22}$\,cm$^{-2}$) &      &  
               &        & & & & & \\
$[1]$ & $[2]$ & $[3]$ & $[4]$ & $[5]$ & $[6]$ & $[7]$ & $[8]$ & $[9]$ & $[10]$ & $[11]$ & $[12]$ & $[13]$ \\ 
\hline

165334.4-414424 & \#41 & all & M1 + M2 & $0.62^{+.23}_{-.23}$ & $0.89^{+.16}_{-.13}$ & 0.94 & 179 & $19.1$ & $21.5$ &  &  & \\
165358.0-415220 & \#171 & all & EPIC   & $0.47^{+.03}_{-.03}$ & $3.42^{+.17}_{-.15}$ & 1.45 & 284 & $4.3$ & $13.7$ &  &  & \\
165404.9-415645 & \#228 & all & M1 + pn & $0.38^{+.07}_{-.06}$ & $3.16^{+.30}_{-.19}$ & 0.97 & 100 & $3.0$ & $7.3$ & 332 & 15.71 & 1.49 \\
165405.9-414509 & \#241 & \#5 & EPIC & $0.37^{+.09}_{-.07}$ & $2.25^{+.19}_{-.16}$ & 0.76 & 80 & $16.1$ & $24.8$ &  &  & \\
165450.6-420201 & \#568 & all & pn & $0.59^{+.21}_{-.15}$ & $3.45^{+.80}_{-.45}$ & 0.87 & 49 & $2.7$ & $10.1$ & [4893] & [16.17] & [2.00] \\
\hline
\end{tabular}
\end{center}
\end{table*}

\begin{table*}
\caption{Same as Table\,\ref{tbl-1}, but for the absorbed 2-T {\tt mekal} models ({\tt wabs*(mekal$_1$+mekal$_2$)}). }\label{tbl-3}
\tiny
\begin{center}
\begin{tabular}{c c c c c c c c c r r c c c} 
\hline
\multicolumn{2}{c}{Source IDs} & Obs.\ & Inst.\ & $N_{\rm H}$            & k$T_1$  & k$T_2$ & $\chi^2_{\nu}$ & d.o.f. & f$_{\rm X}^{\rm obs}$      & f$_{\rm X}^{\rm corr}$ & SBL98 & $V$ & $V-I_C$ \\
XMM\,UJ & Paper~I      &            &        & ($10^{22}$\,cm$^{-2}$) & (keV) & (keV) 
               &        &  & & & & & \\
$[1]$ & $[2]$ & $[3]$ & $[4]$ & $[5]$ & $[6]$ & $[7]$ & $[8]$ & $[9]$ & $[10]$ & $[11]$ & $[12]$ & $[13]$  & $[14]$ \\ 
\hline

165354.6-415303 & \#138 & all & M1 + M2 & $0.20^{+.06}_{-.11}$ & $0.70^{+0.17}_{-0.10}$ & $2.54^{+0.93}_{-0.46}$ & 0.96 & 62 & $3.6$ & $5.1$ &  &  & \\
165359.6-414802 & \#181 & all & EPIC & $0.20^{+.04}_{-.03}$ & $0.63^{+0.05}_{-0.05}$ & $2.48^{+0.27}_{-0.24}$ & 1.23 & 254 & $5.2$ & $7.4$ & [2843] & 14.77 & 1.43 \\
165400.1-414739 & \#187 & all & EPIC & $0.19^{+.03}_{-.04}$ & $0.81^{+0.05}_{-0.07}$ & $2.25^{+0.27}_{-0.22}$ & 0.91 & 173 & $4.2$ & $5.8$ & 285 & 12.94 & 0.57 \\
165401.9-414334 & \#204 & all & M1 & $0.36^{+.17}_{-.08}$ & $0.75^{+0.11}_{-0.09}$ & $3.64^{+1.57}_{-0.83}$ & 0.92 & 56 & $6.7$ & $11.3$ & 304 & 15.64 & 1.45 \\
165406.3-414543 & \#249 & all & EPIC & $0.14^{+.06}_{-.02}$ & $0.87^{+0.10}_{-0.07}$ & $3.39^{+0.69}_{-0.44}$ & 0.92 & 214 & $4.3$ & $5.3$ & 5980 & 16.65 & 1.79 \\
165407.2-415000 & \#254 & \#5 & EPIC & $0.13^{+.13}_{-.06}$ & $0.82^{+0.13}_{-0.13}$ & $3.69^{+0.53}_{-0.80}$ & 0.87 & 69 & $9.6$ & $11.6$ & 356 & 13.29 & 0.82 \\
165407.3-415156 & \#258 & all & EPIC & $0.20^{+.08}_{-.05}$ & $0.76^{+0.09}_{-0.12}$ & $2.58^{+0.38}_{-0.36}$ & 1.05 & 124 & $3.0$ & $4.1$ &  &  &  \\
165408.8-415150 & \#269 & all & EPIC & $0.14^{+.07}_{-.05}$ & $0.74^{+0.10}_{-0.10}$ & $2.82^{+0.33}_{-0.33}$ & 1.11 & 192 & $4.8$ & $6.0$ & 370 & 12.24 & 0.60 \\
165411.5-415004 & \#290 & all & EPIC & $0.08^{+.02}_{-.02}$ & $0.68^{+0.07}_{-0.02}$ & $2.29^{+0.25}_{-0.14}$ & 1.11 & 334 & $6.0$ & $7.1$ & 405 & 15.51 & 1.24 \\
165412.3-415122 & \#297 & all & EPIC & $0.21^{+.04}_{-.04}$ & $0.64^{+0.05}_{-0.05}$ & $2.42^{+0.40}_{-0.29}$ & 1.05 & 131 & $2.7$ & $4.0$ & 413 & 12.75 & 0.49 \\
165413.1-414505 & \#304 & \#2 & EPIC & $0.23^{+.03}_{-.03}$ & $1.31^{+3.12}_{-0.56}$ & $8.2^{+22.5}_{-2.10}$ & 0.99 & 265 & $90.1$ & $105.4$ & 423 & 13.25 & 0.49 \\
165415.1-415019 & \#324 & all & EPIC & $0.20^{+.03}_{-.03}$ & $0.67^{+0.08}_{-0.05}$ & $2.55^{+0.19}_{-0.20}$ & 1.00 & 281 & $9.0$ & $12.3$ & 448 & 15.57 & 1.48 \\
165417.4-414829 & \#350 & all & EPIC & $0.19^{+.05}_{-.05}$ & $0.82^{+0.05}_{-0.07}$ & $2.33^{+0.32}_{-0.19}$ & 0.81 & 168 & $3.7$ & $5.1$ & 472 & 12.42 & 0.75 \\
165423.9-415016 & \#407 & all & EPIC & $0.20^{+.06}_{-.03}$ & $0.80^{+0.06}_{-0.10}$ & $2.37^{+0.30}_{-0.27}$ & 1.30 & 158 & $4.8$ & $6.8$ & 550 & 15.32 & 1.39 \\
165433.5-415632 & \#478 & all & M2 + pn & $0.64^{+.13}_{-.04}$ & $0.63^{+0.08}_{-0.05}$ & $2.37^{+0.50}_{-0.32}$ & 0.91 & 137 & $4.9$ & $15.3$ & 628 & 15.03 & 1.50 \\
165442.5-414959 & \#535 & all & M1 + M2 & $0.69^{+.11}_{-.16}$ & $0.62^{+0.19}_{-0.05}$ & $2.78^{+0.83}_{-0.58}$ & 1.08 & 92 & $4.9$ & $16.3$ & 8459 & 16.64 & 1.62 \\
165445.3-415244 & \#553 & \#2 & EPIC & $0.32^{+.09}_{-.09}$ & $0.91^{+0.30}_{-0.11}$ & $4.54^{+0.94}_{-0.69}$ & 0.86 & 110 & $43.7$ & $57.3$ & 729 & 13.63 & 0.83 \\
165512.2-414630 & \#603 & all & M1 + M2 & $0.35^{+.07}_{-.04}$ & $0.59^{+0.11}_{-0.10}$ & $5.10^{+1.17}_{-0.77}$ & 0.97 & 198 & $25.4$ & $35.7$ & [27130] & [20.65] & [1.36] \\
\hline
\end{tabular}
\end{center}
\end{table*}
\normalsize
\subsection{X-ray light curves}
Light curves were extracted for each source over the same area as for the spectra. They were background subtracted and corrected for the effect of the good time intervals. We used different energy bands and different time bins from 10\,s to 5\,ks and we applied various tests to search for variability (see e.g. Sana et al. \citeyear{HD248}).  Considering the typical count rates of the sources studied here, the most useful time bin to investigate variability is 1\,ks and we thus focus here on the light curves obtained for this binning.

\subsubsection{Comments on individual sources}
\begin{figure}
\begin{center}
\resizebox{8cm}{7.6cm}{\includegraphics{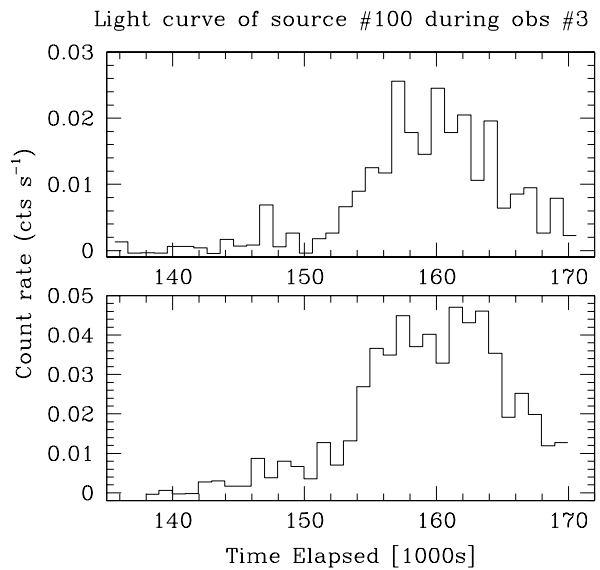}}
\end{center}
\caption{X-ray flare of source \#100 as observed during observation 3 with EPIC-MOS\,2 (top panel) and EPIC-pn (lower panel). The light curve was extracted for pulse invariant channels 500 to 10000 (i.e.\ for energies approximately in the range 0.5 -- 10\,keV). The time is given in ks from the beginning of the first observation.}\label{lc100}
\end{figure}

\begin{figure}
\begin{center}
\resizebox{8cm}{6.43cm}{\includegraphics{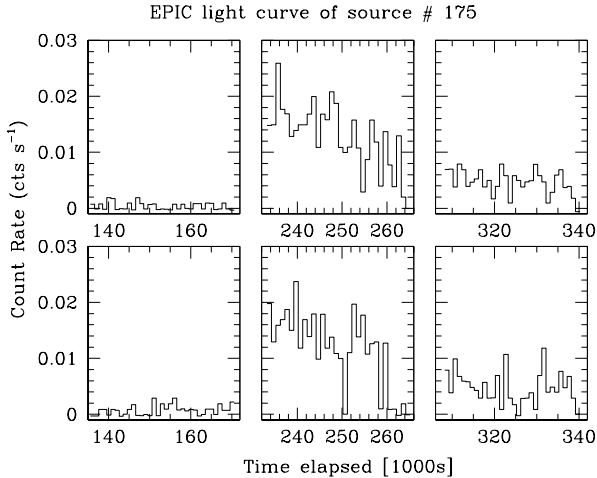}}
\end{center}
\caption{EPIC-MOS\,1 (top panels) and MOS\,2 (lower panels) light curve (for PI in the range 500 to 10000) of source \#175 during {\it XMM} observations 3, 4 and 5 (from left to right). The time is given in ks from the beginning of observation 1.}\label{lc175}
\end{figure}

\begin{figure}
\begin{center}
\resizebox{8cm}{9.24cm}{\includegraphics{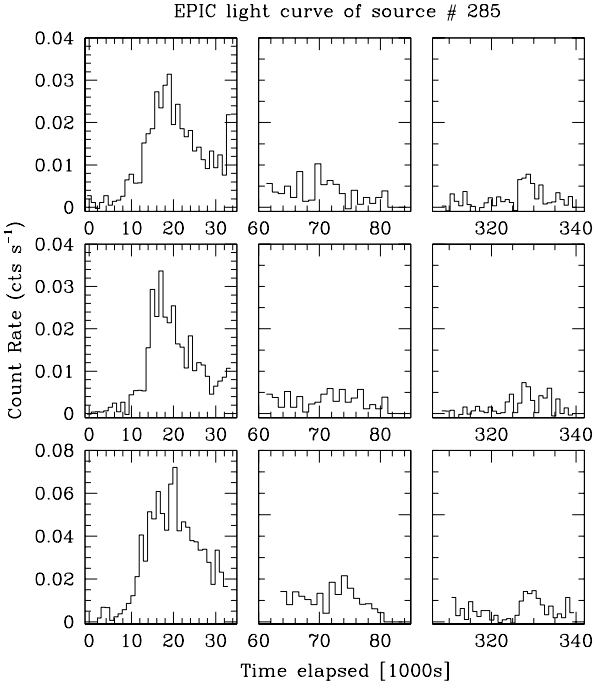}}
\end{center}
\caption{EPIC-MOS\,1 (top panels), MOS\,2 (middle panels) and pn (lower panels) light curve (for PI in the range 500 to 10000) of source \#285 during observations 1, 2 and 5 (from left to right). The time is given in ks from the beginning of observation 1.}\label{lc285}
\end{figure}

\begin{figure}
\begin{center}
\resizebox{8cm}{9.24cm}{\includegraphics{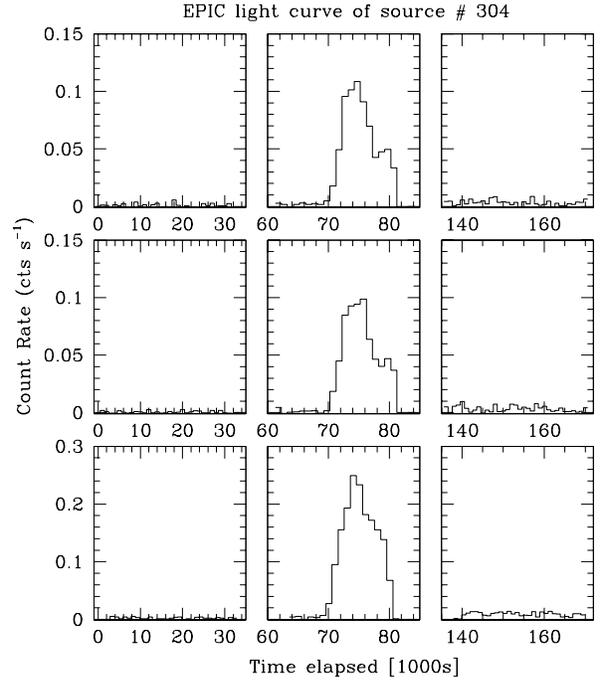}}
\end{center}
\caption{Same as Fig.\,\ref{lc285} but for source \#304 during observations 1, 2 and 3 (from left to right).}\label{lc304}
\end{figure}

\begin{figure}
\begin{center}
\resizebox{8cm}{9.24cm}{\includegraphics{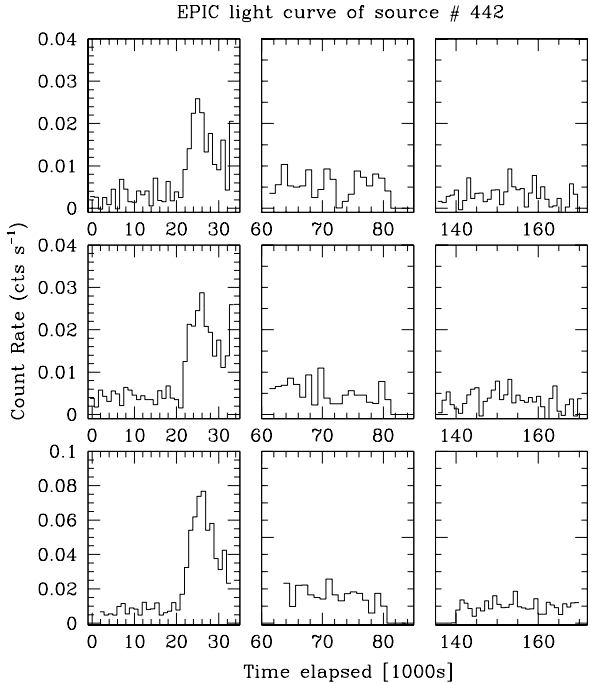}}
\end{center}
\caption{Same as Fig.\,\ref{lc285} but for source \#442 during observations 1, 2 and 3 (from left to right).}\label{lc442}
\end{figure}

\begin{figure}
\begin{center}
\resizebox{8cm}{9.24cm}{\includegraphics{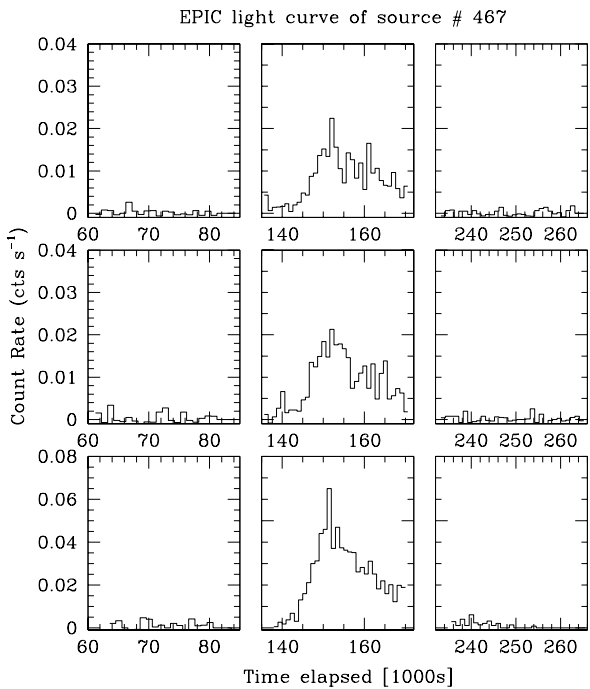}}
\end{center}
\caption{Same as Fig.\,\ref{lc285} but for source \#467 during observations 2, 3 and 4 (from left to right).}\label{lc467}
\end{figure}

\begin{figure}
\begin{center}
\resizebox{8cm}{9.24cm}{\includegraphics{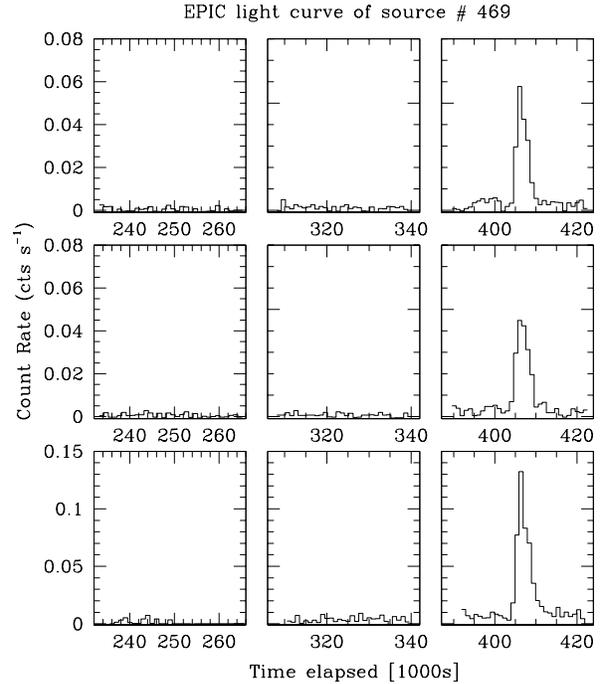}}
\end{center}
\caption{Same as Fig.\,\ref{lc285} but for source \#469 during observations 4, 5 and 6 (from left to right).}\label{lc469}
\end{figure}

\begin{figure}
\begin{center}
\resizebox{8cm}{9.24cm}{\includegraphics{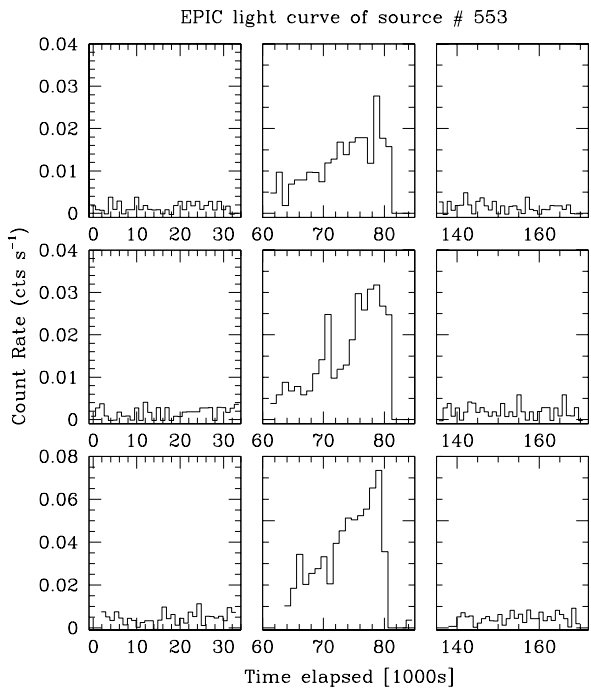}}
\end{center}
\caption{Same as Fig.\,\ref{lc285} but for source \#553 during observations 1, 2 and 3 (from left to right).}\label{lc553}
\end{figure}

\begin{figure}
\begin{center}
\resizebox{8cm}{3.88cm}{\includegraphics{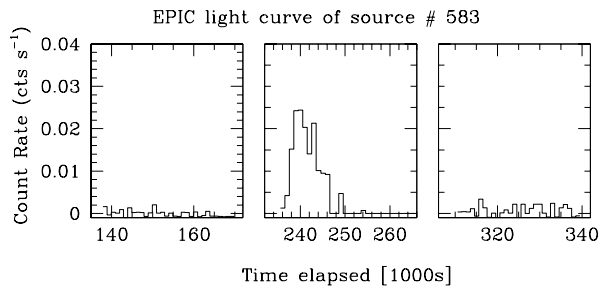}}
\end{center}
\caption{EPIC-pn light curve (for PI in the range 500 to 10000) of source \#583 during observations 3, 4 and 5 (from left to right). The time is given in ks from the beginning of observation 1.}\label{lc583}
\end{figure}

In this section, we discuss the light curves and spectra of X-ray sources that are either variable or have ambiguous spectral fits. Sources \#181, 228, 251, 269, 290, 297, 350, 407, 478, 535, 568 and 603 display no significant variability in their light curve and are not discussed here. Their spectra are however described in Tables\,\ref{tbl-1} to \ref{tbl-3}.
\paragraph{\#41} This source displays little variability. The spectrum is best fitted by an absorbed power law model (Table\,\ref{tbl-2}), although a blackbody model (with k$T = 1.50$\,keV and $N_{\rm H} \leq 0.08 \times 10^{22}$\,cm$^{-2}$) yields a fit that is only slightly poorer ($\chi^2_{\nu} = 0.96$).
\paragraph{\#43} The merged spectrum of this source displays a clear Fe K line. The best quality fit is achieved for a single temperature {\tt mekal} model (see Table\,\ref{tbl-1}). A power law fit yields $\chi^2_{\nu} = 1.08$ for $N_{\rm H} = 0.39 \times 10^{22}$\,cm$^{-2}$ and $\Gamma = 1.65$. The source is somewhat brighter and the spectrum somewhat softer during observation \#6.
\paragraph{\#48} This object displays a flare during observation 4.  
\paragraph{\#100} While no X-ray emission is detected during observations 1, 2, 4 and 6, the source undergoes a strong flare during observation 3 (see Fig.\,\ref{lc100}) and is also clearly detected (though with a much lower count rate) during observation 5. The spectra obtained during observation 3 display a moderate Fe K line and are best fitted with a k$T \sim 4.3$\,keV {\tt mekal} model (see Table\,\ref{tbl-1}), though a power law model with a photon index of 2.2 also yields a $\chi^2_{\nu}$ of 0.90. The spectrum from observation 5 is of much lower quality, but still suggests that the source was much softer (k$T \sim 0.9$\,keV) and about 50 times fainter compared to observation 3.
\paragraph{\#138} The X-ray flux is about 3 times larger during observation 5, though the light curve exhibits no clear flaring behaviour.
\paragraph{\#171} No significant variability is found for this source. The spectrum cannot be fitted by a single temperature {\tt mekal} model, though a 2-T model ($N_{\rm H} = 0.79 \times 10^{22}$\,cm$^{-2}$, k$T_1 = 0.27$\,keV and k$T_2 = 1.61$\,keV) yields $\chi_{\nu}^2 = 1.30$, slightly better than the power law model quoted in Table\,\ref{tbl-2}.
\paragraph{\#175} This source is clearly detected in all our observations. During the first three pointings, the average MOS count rate is about $6 \times 10^{-4}$\,cts\,s$^{-1}$. At some point between observation 3 and 4, the X-ray flux of the source has increased by at least a factor 30 (see Fig.\,\ref{lc175}). Indeed, during the fourth pointing, the count rate slowly decreases from $\sim 2.0 \times 10^{-2}$ to $\sim 0.6 \times 10^{-2}$\,cts\,s$^{-1}$. During the last two observations, the source is still brighter than at the beginning of our campaign though the average MOS count rate decreases slowly from $\sim 5.1 \times 10^{-3}$ to $\sim 2.6 \times 10^{-3}$\,cts\,s$^{-1}$. The spectrum of observation 4 indicates a rather hot (k$T = 4.3$\,keV, see Table\,\ref{tbl-1}) plasma, though a power law fit ($\Gamma \sim 2.0$) is only marginally poorer ($\chi^2_{\nu} = 1.05$). During observations 5 and 6, the spectrum is slightly softer (k$T$ decreases to $3.4$\,keV or $\Gamma$ increases to 2.6).
\paragraph{\#204} A flare is seen in the MOS\,1 count rate towards the end of observation 4. However, no pn or MOS\,2 data are available for this source to confirm this result.
\paragraph{\#241} The count rate of this source is about a factor 15 (respectively 6) larger during observation 5 (resp.\ 6) than during observations 1 to 4. It is likely that the source has undergone a flare at some time between the end of the fourth and the beginning of the fifth pointing, and that our data actually cover the decreasing tail of this flare. 
\paragraph{\#249} No significant variability is found for this source during the different pointings, although the EPIC count rates are somewhat larger during the last observation.
\paragraph{\#254} During observation 5, the source exhibits a small flare with an increase of the count rate by a factor 5 to 6 which is consistently seen in all three EPIC instruments.
\paragraph{\#258} The X-ray flux is largest during observation 1 and decreases slowly afterwards. 
\paragraph{\#285} This source experiences a strong flare during the first observation and a subsequent decrease of the count rate during the second observation (see Fig.\,\ref{lc285}). A small flare is seen in all EPIC instruments during the fifth pointing. The spectrum obtained during the flare of the first observation can be fitted by a 1-T {\tt mekal} model (see Table\,\ref{tbl-1}), although a fit of equal quality can be obtained for a power law model with $N_{\rm H} = 0.37 \times 10^{22}$\,cm$^{-2}$ and $\Gamma = 2.10$.
\paragraph{\#304} A strong flare is seen during the second observation where the count rates increase by about a factor 100 (Fig.\,\ref{lc304}). The spectrum during the flare exhibits a prominent Fe K line. 
\paragraph{\#324} The source is about a factor 2 brighter during the fourth observation, whilst the count rates of the different instruments during the other observations exhibit no coherent significant variability. 
\paragraph{\#442} A flare with an increase of the count rate by a factor 10 occurs during the first observation. During the subsequent observations, the count rate slowly decreases (Fig.\,\ref{lc442}). 
\paragraph{\#467} The source exhibits a flare during the third observation (Fig.\,\ref{lc467}). During the flare, the spectrum is equally well described by a k$T = 5.7$\,keV single temperature {\tt mekal} model (see Table\,\ref{tbl-1}) or a power law model with $\Gamma = 1.81$ and $N_{\rm H} = 0.34 \times 10^{22}$\,cm$^{-2}$.
\paragraph{\#469} Another flaring source with an increase of the count rate by a factor $\sim 60$ during the last observation (Fig.\,\ref{lc469}). During the flare, the spectrum is again equally well described by a single temperature {\tt mekal} model (see Table\,\ref{tbl-1}) or a power law model with $\Gamma = 2.09$ and $N_{\rm H} = 0.38 \times 10^{22}$\,cm$^{-2}$.
\paragraph{\#553} This source displays a steady increase of its count rate over the duration of the second observation (Fig.\,\ref{lc553}). The light curve is truncated at the end of the observation due to the rejection of a soft proton flare from our data (see Paper~I). During the remaining observations, little variability is seen. While we adopt the 2-T {\tt mekal} model (see Table\,\ref{tbl-3}), which fits the Fe K line better, we note that a power law model with $\Gamma = 2.27$ and $N_{\rm H} = 0.42 \times 10^{22}$\,cm$^{-2}$ yields a fit of formally equal quality.
\paragraph{\#583} A flare is seen during the fourth observation (Fig.\,\ref{lc583}). The spectrum has a rather low quality because of a low actual count rate. 

\begin{table*}
\caption{Properties of the flaring sources: $\tau$ yields the $1/e$ decay time, k$T_{\rm obs}$ is the best fit plasma temperature during the flare (a value between brackets corresponds to the hottest component of a 2-T fit) and $l$ is the loop half-length. For comparison, the stellar radii computed from the effective temperatures and bolometric magnitudes derived in Sect.\,\ref{optcoun} are provided in the sixth column. Column 7 indicates the value of the $R - H\alpha$ index compared to the value of this index for main sequence stars with the same intrinsic $V - I_C$ (whenever available). Finally, the last two entries correspond to the $JHK$ colours derived from the 2MASS measurements. }\label{tbl-4}
\begin{center}
\begin{tabular}{r c r c c c c c c} 
\hline
Source & Obs. & \multicolumn{1}{c}{$\tau$} & k$T_{\rm obs}$ & $l$ & $R_*$       & $\Delta(R - H\alpha)$ & $J - H$ & $H - K$ \\ 
       &     & \multicolumn{1}{c}{(ks)} &(keV)   & (R$_{\odot}$) & (R$_{\odot})$ &                      &         & \\
\hline
 \#48 & 4         & 8.8 & 3.43   & 1.0 &     &         & 0.641 & 0.215 \\ 
\#100 & 3         & 5.3 & 4.32   & 0.7 & 1.3 &         &       &       \\
\#204 & 4         &     & (3.64) &     & 2.6 & \hspace*{2.5mm}0.03 & 0.620 & 0.171 \\
\#254 & 5         & 7.3 & (3.69) &     & 3.7 & \hspace*{2.5mm}0.08 & 0.388 & 0.067 \\
\#285 & 1         & 9.4 & 4.44   & 1.2 &     &         & 0.788 & 0.343 \\
\#304 & 2         & 5.5 & (8.2)  &     & 1.6 &         & 0.204 & 0.071 \\
\#442 & 1         & 7.6 & 4.04   & 0.9 & 0.8 &         &       &       \\
\#467 & 3         &16.0 & 5.70   & 2.4 & 1.7 & \hspace*{2.5mm}0.45 & 0.991 & 0.432 \\
\#469 & 6         & 2.1 & 4.17   & 0.3 & 2.1 & $-0.04$ &       &       \\
\#583 & 4         & 5.2 & 5.58   & 0.8 &     &         & 0.460 & 0.110 \\
\hline
\end{tabular}
\end{center}
\end{table*}
\subsubsection{X-ray flares}
Most of the flares displayed in Figs.\,\ref{lc100}-\ref{lc583} exhibit an impulsive rise of the source X-ray luminosity within a few ks followed by a slower decay.  The exceptions are sources \#100 and probably \#553 that display a much slower rise on timescales of 10\,ks and $\geq 20$\,ks respectively. Other sources (such as \#467) show a slower flux increase than expected for typical flares.  Stelzer et al. (\citeyear{Stelzer}) suggested that such a shape could be explained by  a rotational modulation of the X-ray emission produced by the flare. In such a configuration, the flare actually occurs on the hidden face of the star and then becomes slowly visible with the rotation of the object. It is however difficult to investigate more this possibility in the present case as the rotational periods of our objects are not known. We finally note that the detected flaring sources display a harder spectrum than quieter sources, with an average temperature during flares about 4-4.5~keV.

Some quantitative insight into the flare events can be obtained assuming that the flaring plasma is confined in a closed coronal loop as is the case for solar flares. For these situations, Serio et al.\ (\citeyear{Serio}) have established a simple analytical relation between the loop half-length $l$, the maximum temperature at the top of the loop $T_{\rm max}$ and the thermodynamic decay time $\tau$:
$$l \sim 0.129\,\tau\,\sqrt{\mathrm{k}T_{\rm max}}$$
where $l$ is expressed in R$_{\odot}$, k$T$ in keV and $\tau$ in ks. This relation is valid for the initial decay phase during which the emission measure decay is roughly exponential and has been applied to flares in a variety of active stars (e.g.\ Briggs \& Pye \citeyear{Briggs}, Giardino et al.\ \citeyear{Giardino} and Favata \citeyear{Favata}). When the temperatures are determined from spectral fits of EPIC data, a correction has to be applied to the observed temperature to derive the maximum temperature at the top of the loop (see Giardino et al.\ \citeyear{Giardino}, Briggs \& Pye \citeyear{Briggs}).

Whenever possible, we have determined the $1/e$ decay times of the flares by a least square fit of an exponential decay to the decay phase of the observed light curves. Usually, the results for the different EPIC instruments are in good agreement (to better than 10\% of $\tau$), except for sources \#254, \#285, \#442 for which we have a larger dispersion of $\sim 20$\% between the values determined from the three light curves. For sources \#175 and \#241, our data do not cover the maximum of the flare and are hence not suited to constrain the decay time, although we note that the light curve of \#241 during observations 5 and 6 is well fitted with a single exponential decay suggesting a rather large decay time ($\geq 30$\,ks).  

Table\,\ref{tbl-4} yields the properties of the flaring X-ray sources with a sufficiently large count rate to study their spectrum and light curve. All flaring sources have rather large temperatures during the flare. For those spectra that were fitted with a 2-T model, Table\,\ref{tbl-4} indicates the temperature of the hotter model component between brackets. The decay times range from about 2\,ks to 16\,ks and the loop half-lengths are in the range 0.3 to 2.4\,R$_{\odot}$. Most loops appear rather compact ($l \leq R_*$) except perhaps for source \#467. 

Favata (\citeyear{Favata}) notes that while most flaring structures are indeed relatively compact, extremely slowly decaying flares have been detected in some young stellar objects (with $\tau \sim 40$\,ks) which correspond to rather large loops ($l \sim 14$\,R$_{\odot}$). Favata (\citeyear{Favata}) accordingly suggests that the magnetic structures that produce these large loops link the star to the circumstellar disk. We note that, although our dataset is not suited to clearly identify slowly decaying flares, the decay seen for source \#241 might be a candidate for such an event. Unfortunately, the source could not be associated with an optical counterpart though a near-IR counterpart was found. Hence, we can not exclude that \#241 is a background source.

In this context, it is also interesting to note that among the five objects in Table\,\ref{tbl-4} for which a $\Delta(R - H\alpha)$ value can be computed, only source \#467 shows a strong H$\alpha$ emission. Further evidence for circumstellar material around this source comes from its near-IR excess. From Table\,\ref{tbl-4}, it appears that \#467 is also the source with the largest loop half-length ($l \sim 1.4\,R_*$) in our sample.

\subsection{X-ray fluxes}
\begin{figure}
\begin{center}
\resizebox{8cm}{7.8cm}{\includegraphics{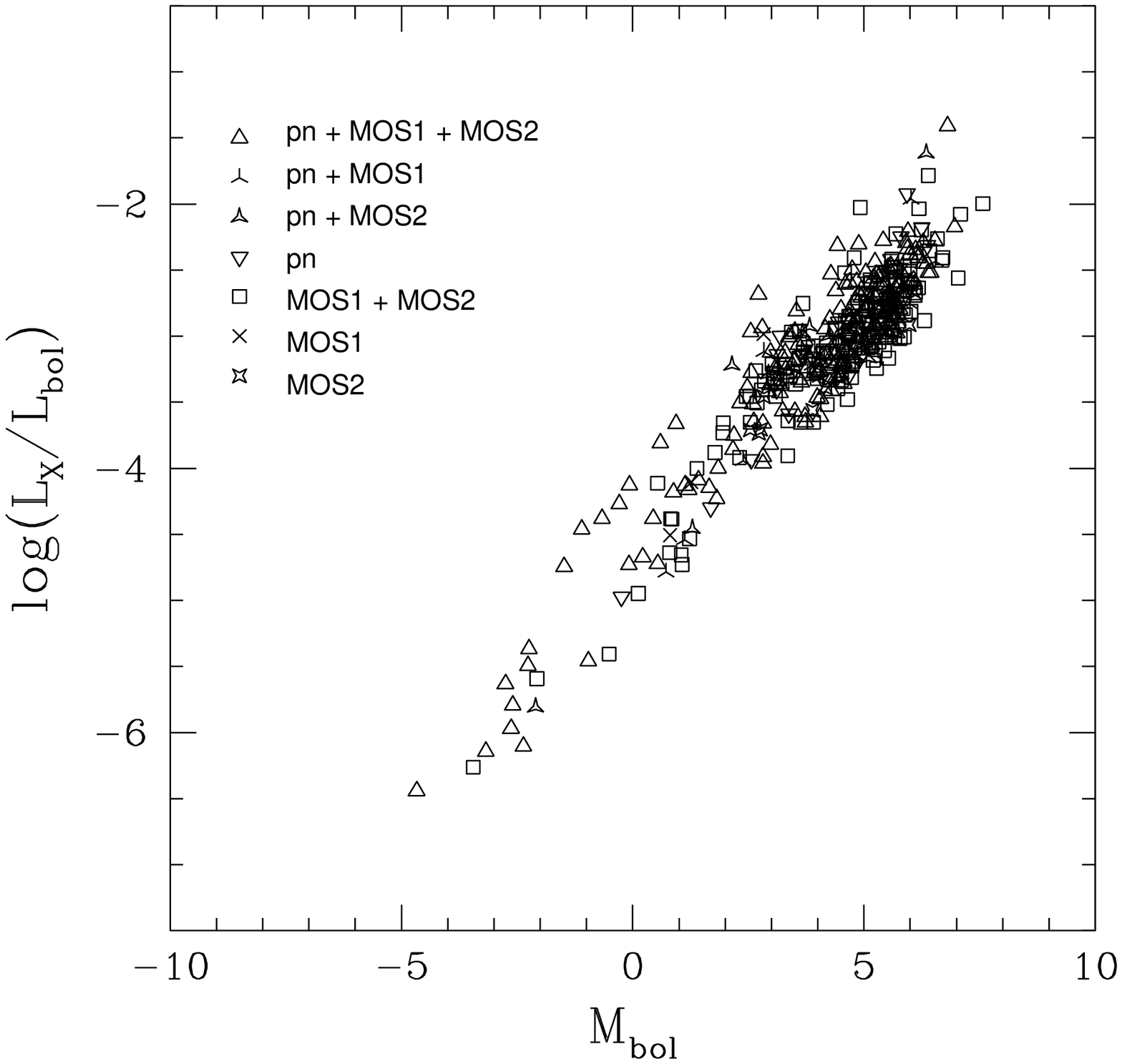}}
\resizebox{8cm}{7.8cm}{\includegraphics{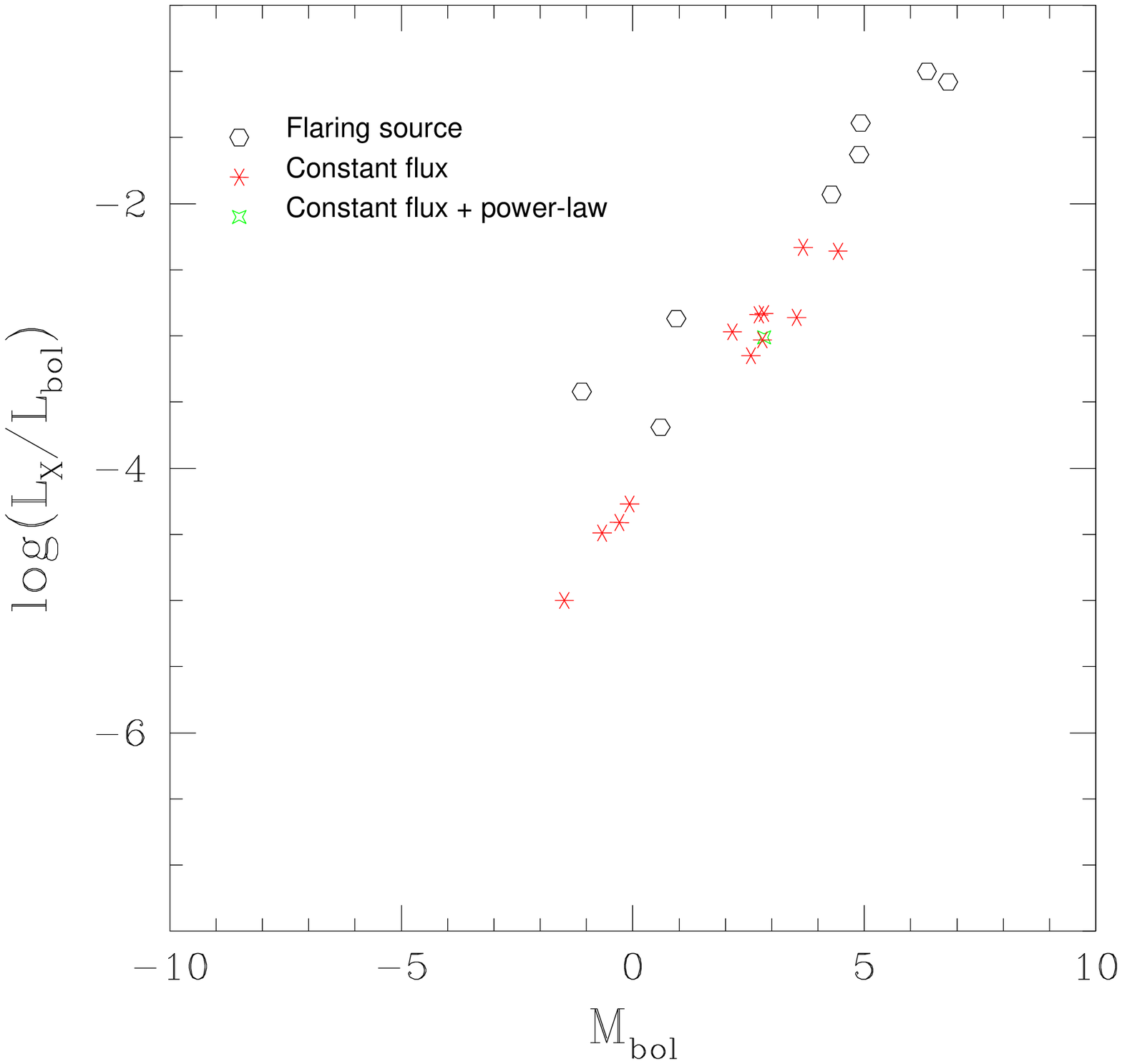}}
\end{center}
\caption{{\bf Top panel:} logarithm of $L_{\rm X}/L_{\rm bol}$ as a function of the bolometric magnitude for the EPIC sources with optical counterparts in the SBL98 catalogue (see text). $L_{\rm X}$ refers to the X-ray luminosity in the energy range 0.5 to 10\,keV. The various symbols correspond to different combinations of EPIC instruments used in the determination of the X-ray luminosity (see the insert in the upper left corner). {\bf Bottom panel}: $\log{L_{\rm X}/L_{\rm bol}}$ as a function of the bolometric magnitude for those EPIC sources with optical counterparts in the SBL98 catalogue and for which a detailed modelling of the spectra was possible (see Tables \ref{tbl-1}, \ref{tbl-2} and \ref{tbl-3}). Asterisks correspond to sources that have a rather constant flux during our six observations, whereas hexagons stand for flaring sources. }\label{lxlbol}
\end{figure}

Using the bolometric magnitudes derived above from the SBL98 photometry, and the vignetting and background corrected EPIC count rates determined by the source detection algorithm, we can build an empirical $L_{\rm X}/L_{\rm bol}$ relation for the sources in the NGC\,6231 field. In order to convert the EPIC count rates into X-ray luminosities in the 0.5 -- 10\,keV range, we took the average of the available count rates of the pn, MOS1 and MOS2 instruments assuming a conversion factor 2 between the MOS and pn count rates. Next, we computed a flux/pn count rate conversion factor assuming that the sources have an absorbed optically thin thermal plasma spectrum with a temperature of 1\,keV and $N_{\rm H} = 5.8 \times 10^{21} \times \overline{E(B-V)} = 2.7 \times 10^{21}$\,cm$^{-2}$ (Bohlin et al.\ \citeyear{Bohlin}). Note that adopting a temperature of 2\,keV (more typical of the spectra of the brightest sources) would only slightly increase this conversion factor (by 14\%). Finally, we converted the unabsorbed fluxes into luminosities, assuming that all sources are members of NGC\,6231 ($DM = 11.0$). 
Figure\,\ref{lxlbol} displays the resulting $\log{L_{\rm X}/L_{\rm bol}}$ as a function of $M_{\rm bol}$. This value increases from about $-7$ for B0 stars (the hottest objects plotted in this figure) to almost $-2$ for the lower mass objects. A similar trend was already observed in NGC\,6530 (Rauw et al.\ \citeyear{M8}).

Of course the results in Fig.\,\ref{lxlbol} could be somewhat biased by the fact that some of the sources are detected only during their flaring activity. We have repeated the same procedure for the brighter sources discussed hereabove. In the bottom panel of Fig.\,\ref{lxlbol}, we distinguish between the flaring and non flaring sources. We find that the average luminosity during a flare can reach up to $3 \times 10^{32}$\,erg\,s$^{-1}$ (source \#304) and $\log{L_{\rm X}/L_{\rm bol}}$ can temporarily increase to values as large as $-1.00$ (source \#100). On the other hand, the brighter steady sources also display rather large X-ray luminosities, up to $\sim 4 \times 10^{31}$\,erg\,s$^{-1}$, as well as large $\log{L_{\rm X}/L_{\rm bol}}$, up to $\sim -2.4$. Therefore, we conclude that even non-flaring PMS sources in NGC\,6231 can be rather X-ray bright.
\section{Discussion}
Our {\it XMM-Newton} campaign provides the deepest X-ray view ever of NGC\,6231. While the X-ray images of the cluster are dominated by its high-mass members (see Paper~I), we detect a substantial number of sources that have faint optical counterparts. The typical spectra of these sources, their flaring behaviour, their X-ray luminosities as well as the location of their optical counterparts in the H-R diagram indicate that these sources form a population of PMS stars. Compared to previous optical investigations of the cluster (Sung et al.\ \citeyear{SBL98}), the number of PMS stars is increased by a factor 20 at least. 

An interesting feature is the age spread among the low-mass PMS objects ($\sim 11$\,Myr) that exceeds the main sequence lifetime of the massive O stars in the cluster. This feature was already noted by previous authors. Raboud et al.\ (\citeyear{RCB}) obtained photometry of NGC\,6231 in the Geneva system. Fitting isochrones to the main sequence and post-main sequence stars believed to be single, these authors derived an age between 3.2 and 4.5\,Myr for the more massive stars, whilst they noted that the PMS stars detected in their observations were mostly older than 3.2\,Myr and had ages up to 10\,Myr. A similar situation was found by Sung et al.\ (\citeyear{SBL98}) who derived an age of 2.5 -- 4\,Myr for the massive stars in the cluster, while they found an age spread of 11\,Myr in the PMS objects. Raboud et al.\ (\citeyear{RCB}) accordingly suggested that there was a continuous low-mass star formation activity in NGC\,6231 at previous times, which came to an end about 3.2 -- 4.5\,Myr ago when the massive O stars of the cluster formed during a nearly coeval burst. From our results, the age distribution of X-ray selected low-mass stars peaks around 1 -- 4\,Myr, i.e.\ at the time of the formation of the early-type cluster members. However, the formation of low mass stars in NGC\,6231 started already more than 10\,Myr ago at a rather slow rate. The star formation rate subsequently increased, culminating in a starburst event during which most of the PMS as well as the more massive objects formed. Then, about 1\,Myr ago, the formation of stars in NGC\,6231 came to an end rather abruptly. 

\begin{figure}
\begin{center}
\resizebox{8cm}{8.0cm}{\includegraphics{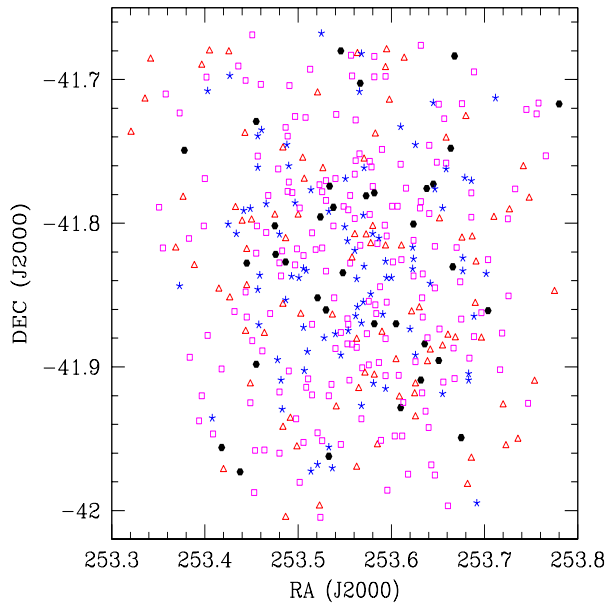}}
\end{center}
\caption{Spatial distribution of the X-ray selected low-mass stars in NGC\,6231. The triangles, squares, stars and filled hexagons indicate stars with ages of less than 2\,Myr, between 2 and 5\,Myr, between 5 and 9\,Myr and more than 9\,Myr respectively. }\label{carte}
\end{figure}

In the case of NGC\,6530, Damiani et al.\ (\citeyear{Damiani}) report on a spatial age gradient, suggesting a sequence of star formation events. To check whether a similar situation holds for NGC\,6231, we have plotted the spatial distribution of X-ray selected PMS stars of different ages in Fig.\,\ref{carte}. It can be seen that, in the case of NGC\,6231, there exists no clear separation between the different groups. While this could be a projection effect (if star formation were progressing along our line of sight towards the cluster), we nevertheless conclude that there is no evidence in our data for a spatial age gradient in NGC\,6231.

The star formation history of NGC\,6231 provides constraints on future modelling of massive star formation in open clusters. Several properties of the NGC\,6231 cluster could be explained by a massive star formation scenario involving physical collisions in a dense cluster core (Bonnell et al.\ \citeyear{BBZ}). These features are the large binary fraction among the massive stars in this cluster, the spatial concentration of PMS stars around the massive cluster members and last but not least, the age difference between the PMS objects and the more massive stars. Indeed, N-body simulations of compact cluster cores including the effects of stellar evolution and physical collisions (Portegies Zwart et al.\ \citeyear{PZ}) show that, because of mass segregation, physical collisions between stars are frequent in the cluster core. The resulting mergers could lead to the formation of very massive ($\geq 100$\,M$_{\odot}$) stars within 3 -- 4\,Myr. The stars are rejuvenated by each new collision, therefore considerably extending their lifetime. The products of such a process would thus appear as some sort of `blue stragglers' in the H-R diagram of the cluster.

Alternatively, a moderate age difference could also result if massive star formation proceeds through accretion as proposed by Norberg \& Maeder (\citeyear{Norberg}) and Behrend \& Maeder (\citeyear{BM1}). These authors assumed that the PMS objects accrete a constant fraction of the observed outflow rates, which in turn increases with the stellar luminosity and hence the stellar mass. In their models, the stars become visible on the birthline once a significant part of the surrounding cocoon has been dispersed. For instance, in the models of Norberg \& Maeder (\citeyear{Norberg}), an accreting protostar that will eventually form a 120\,M$_{\odot}$ star would spent half of the accretion time below a mass of about 2\,M$_{\odot}$. The time needed for an accreting protostar to increase its mass from 2 to 120\,M$_{\odot}$ would then be about 1\,Myr. These models require rather high accretion rates (of order $8 \times 10^{-3}$\,M$_{\odot}$\,yr$^{-1}$ to form an 85\,M$_{\odot}$ star, Behrend \& Maeder \citeyear{BM1}). Though this value seems high, it is apparently not forbidden by the results of Wolfire \& Cassinelli (\citeyear{Wolfire}). The models of Behrend \& Maeder (\citeyear{BM1}) accordingly predict that low mass stars form first and high mass stars form slightly later. However, the duration for the formation of an 80\,M$_{\odot}$ star is only about 10\% larger than that of an 8\,M$_{\odot}$ star. Henceforth, this scenario alone cannot explain the large age spread observed in Fig.\,\ref{ages}.

\section{Conclusions}

In this third paper of the series, we discussed the properties of the NGC~6231 X-ray sources with faint optical counterparts. Their typical X-ray spectra and luminosities, their flaring behaviour  as well as their location in the H-R diagram indicate that the bulk of these objects are most probably PMS stars with masses between 0.3 and 3.0~\,M$_{\odot}$. Their observed age distribution suggests that the low-mass stars in NGC 6231 started their formation at a relatively slow rate over 10~Myr ago. This process then culminated 1 to 4 Myr ago in a starburst-like event that coincided with the time of formation of the massive stars in the cluster. No difference in the location of the cTTs and wTTs in the cluster as well as no spatial age gradient were observed.

The X-ray spectra of the brightest PMS sources are well reproduced by rather hard thermal plasma models. About one third of them displays flaring activity during the {\it XMM-Newton} campaign. The observed $\log{L_{\rm X}/L_{\rm bol}}$ values increase with decreasing bolometric luminosity up to values of $\log{L_{\rm X}/L_{\rm bol}} \sim -2.4$ for non-flaring sources and to even larger values during flares.

Finally, we qualitatively discussed possible implications for the massive star formation in NGC 6231. Briefly a scenario implying the merging of low-mass objects to create high-mass stars would probably allow to explain most of the observational properties of the cluster. In particular, it will likely be able to reproduce the large binary fraction seen in the massive star population as well as the important time (several Myr) separating the beginning of formation of the low- and high-mass stars. 

To conclude, we outline that X-ray emission is probably one of the best selection criteria for PMS objects. Beside a detailed understanding of the massive star population, we now have probably identified a significant fraction of the PMS population of the cluster.  The determination of the initial mass function of both the high-mass and low-mass star populations could bring further constraints on the star formation in NGC 6231 and, more generally, on the possible link between the formation of these two populations. This task is deferred to a forthcoming paper in the series.

\section*{Acknowledgments}

The Li\`ege team is greatly indebted to the Fonds National de la Recherche Scientifique (Belgium) for multiple assistance. This research is also supported in part by contract P5/36 ``P\^ole d'Attraction Interuniversitaire'' (Belgian Federal Science Policy Office) and by PRODEX contracts related to the {\it XMM-Newton} and {\it INTEGRAL} missions. H. Sung acknowledges the support of the Korea Science and Engineering Foundation (KOSEF) to the Astrophysical Research Center for
the Structure and Evolution of the Cosmos (ARCSEC$''$) at Sejong University.  The SIMBAD database has been consulted for the bibliography.

\bsp

\label{lastpage}

\end{document}